NOTE: The version of record of this article, first published in Nature Human Behaviour, is available online at Publisher’s website: https://doi.org/10.1038/s41562-023-01677-8

**TITLE: Some scientific knowledge is a dangerous thing: overconfidence grows non-linearly with knowledge**.

Simone Lackner[1]†, Frederico Francisco[3]†, Cristina Mendonça[1]†, André Mata[4], Joana Gonçalves-Sá[1]‡,[2]*

1. Laboratório de Instrumentação e Física Experimental de Partículas, Lisboa, Portugal
2. Nova School of Business and Economics, Carcavelos, Portugal
3. Departamento de Física e Astronomia, Faculdade de Ciências, Universidade do Porto, Portugal
4. Faculdade de Psicologia, Universidade de Lisboa, Portugal

* Corresponding author (joanagsa@lip.pt)
+ These authors contributed equally to this work
‡ Current address

## ABSTRACT
Overconfidence is a prevalent problem and particularly consequential in its relation with scientific knowledge: being unaware of one's own ignorance can affect behaviours and threaten public policies and health. We introduce both analytical and methodological changes to the study of confidence in science knowledge and in attitudes towards science and apply them to four large surveys, spanning 30 years in Europe and the USA. We propose a new indirect confidence metric that does not rely on self-reporting or peer comparison, and study how knowledge and confidence vary across their full scale. We find that confidence grows much faster than knowledge, giving rise to a non-linear relationship, with the largest confidence gaps appearing at intermediate knowledge levels. These high-confidence / intermediate-knowledge groups also display the least positive attitudes towards science, with important consequences for science communication.
These results are contrary to current models, including the predictions of the Dunning-Kruger effect, and we discuss how our model, if correct, can guide future research and communication policies.

## INTRODUCTION
It has been argued that "no problem in judgment and decision making is more prevalent and more potentially catastrophic than overconfidence" [1]. Overconfidence can be broadly defined as a bias that makes people have a subjective assessment of their own aptitude that is greater than the objective accuracy of such aptitude. These subjective assessments lead to calibration errors with people both overestimating (overconfidence) or underestimating (underconfidence) their ability. A well-studied example concerns how confidence varies with knowledge: does possessing knowledge also come with accurate metacognition about one's knowledge? That is, are the ones who know less aware of it or, conversely "Do those who know more also know more about how much they know"[2]? If there were perfect metacognition, one should expect a linear relationship between how much one knows (Knowledge) and how much one thinks one knows (Confidence) (Figure 1A, yellow line). However, there is now general agreement that this perfect linear relationship does not exist and that miscalibrations in the internal representation of accuracy can have dire consequences [3-6]. Therefore, accurately identifying populations more at risk of overconfidence is fundamental. Notably, Dunning and Kruger [7] have shown that, at best, there is a weak correlation between knowledge and confidence (Figure 1A, green dashed line), with the least knowledgeable being more likely to overestimate their skills. In particular, the Dunning-Kruger Effect (DKE) has been identified in controversial anti-science movements, such as vaccine hesitancy and opposition to genetically modified foods [8,9], and overconfidence can play a role in pandemic control and science communication in general.

**Figure 1**

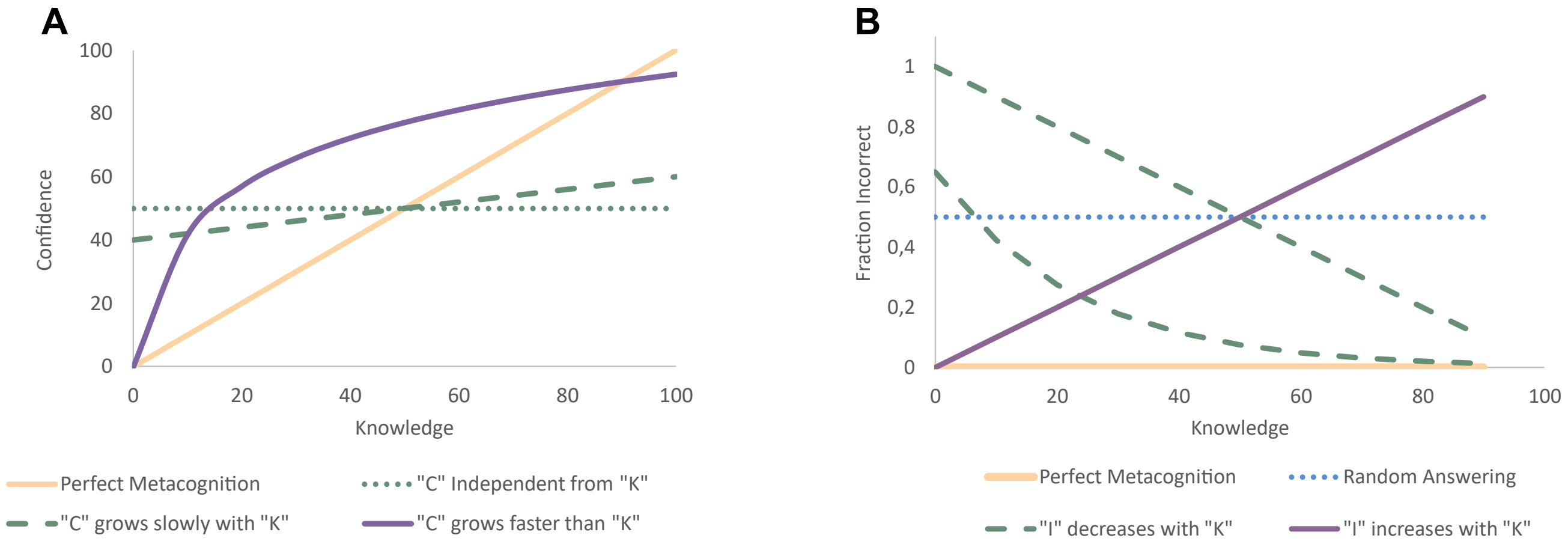

**Fig. 1. Possible relationships between confidence and knowledge**. **(A)** Comparison of different theoretical models: Perfect metacognition would predict a linear relationship between knowledge and confidence (yellow line, the more one knows, the more one is confident about one's knowledge); the Dunning-Kruger effect (DKE; poor or no correlation between knowledge and confidence - dashed green line); independence (dotted green line); an example of confidence growing non-linearly and faster than knowledge (solid purple). **(B)** Comparison of expected patterns of average normalized incorrect (I) to "Don't Know" (DK) ratios assuming different response models. X-axis goes from zero correct answers (all answers are either I or DK) to all correct answers (zero I and zero DK): perfect calibration (yellow line, respondents never answer incorrect), DKE (green dashed lines, respondents are more likely to answer wrongly at low knowledge levels), random answering (blue dotted line, respondents are as likely to answer I as DK); confidence grows faster than knowledge (solid purple line).

Since the publication of the original DKE, the field has accumulated a large body of evidence and evolved to suggest a more nuanced pattern than the original DKE proposed. For example, several studies indicate that the main tenet of the effect - that the unskilled are unaware of their lack of skill–, might not be universal [10-13], such that not all unskilled are unaware, and different authors, including Dunning, have tried to replicate the effect in different circumstances and countries, finding that overconfidence follows individual and cultural trends [14]. Moreover, gauging confidence or knowledge faces its own challenges as the metrics themselves are never perfect and might not be universal or even independent. Therefore, there is reason to believe that the relationship between confidence and knowledge is modulated by several factors [15-17], including methodological as well as analytical issues, among which we highlight two. First, the confidence measures typically used in this research often rely on self-reported metrics, whereby participants are asked how well they believe they performed on a given task, and this can introduce important biases: it is well-established that study participants often show social desirability bias [18,19] and struggle to accept limitations publicly [20,21], with many people offering answers to survey questions even when the subject is fictitious [22]. Furthermore, people are often motivated to view themselves in a positive light, which can also lead to (positive) distortions in self-report [23]. More subtle measures of confidence (e.g., response times, skin conductance, brain imaging) suggest that the unskilled often show doubt at a more implicit level [24]. Second, Dunning-Kruger-type studies typically present both knowledge and confidence in comparative scales, in which respondents are asked to compare their performance to the performance of others (e.g., student participants in [8] were asked to compare their ability to recognize humour in relation to the average student, using percentile ranking), making it difficult to distinguish between poor self-assessment or underestimation of others' abilities. In addition, these comparative assessments are often presented in quartiles, which can hide important variability, by grouping together respondents with very different mean knowledge or confidence, as Dunning and Kruger have also observed [25]. However, as the extremes of the distributions often include very little data, not grouping them could also lead to treating noise as signal. Still, it is broadly accepted that the least knowledgeable tend to show the biggest confidence gap or, in other words, that the least knowledgeable are often the more overconfident, and this was recently re-stated in relation to controversial science issues, in a study that assesses both objective scientific knowledge and subjective self-assessment [51].

Here, we employ a different methodological and analytical approach to address both these issues. First, we introduce a new metric, based on the premise that "Don't Know" versus incorrect answers to knowledge questionnaires can be used as a proxy for confidence, and we examine how this indirect confidence varies with knowledge. Second, we apply our metric to several large surveys, conducted over 30 years in Europe and the USA. Third, we introduce a new analytical approach and analyse these surveys over their full range of variability (instead of considering only performance quartiles) and compare their results to three different models: the already described metacognition model expectation that confidence should grow linearly with knowledge, the Dunning-Kruger pattern of almost no relationship between the two variables, and a third model that introduces the possibility of respondents guessing, which could lead to an overestimation of individual knowledge.

We find that, contrary to previous work, and unlike the DKE, overconfidence is not highest among the least knowledgeable, but rather at intermediate knowledge levels, in all three models. We further test this result in a specifically designed survey that includes a more common confidence metric, and confirm the trend that confidence grows faster than knowledge, leading to populations that have some knowledge but strongly overestimate it. Finally, we investigate public attitudes towards science and find that this intermediate knowledge group also corresponds to the one displaying the most negative attitudes. We discuss the impact of our findings in the broad field of overconfidence studies and in the context of science communication.

## RESULTS

### Non-linear relationship between knowledge and confidence

Figure 1A depicts possible models for the relationship between knowledge and confidence: in the case of perfect metacognition, confidence should grow linearly with knowledge (the more one knows, the more confident one should be about one's knowledge – yellow line); if confidence is independent from knowledge, we should observe no correlation (horizontal green dotted line); and if confidence grows faster than knowledge, we could observe a non-linear relationship between the two variables (example in the solid purple line). In the case of the Dunning-Kruger effect (DKE) confidence would either grow moderately with knowledge (dashed green line) or be largely independent (the "better than average effect" would also predict a close to horizontal line, with higher than 50% average confidence).

As mentioned, most studies that assess this relationship use different types of tasks (from grammar tests to humor) but mostly rely on self-reported confidence measures, which may introduce bias and confound the relationship between metacognition and motivated processes. Moreover, these results are often presented in quartiles, hiding important variance, particularly in the most extreme groups, where different knowledge and confidence levels might be treated as one. To minimize these issues, we introduce an indirect confidence variable (Figure 1B) and apply it to very large surveys.

The indirect confidence metric is conceptually very simple, relies on widely consensual science knowledge, and can be applied to all knowledge questionnaires of the format True/False/Don't Know: when given those three options, an individual with perfect metacognition should either offer the correct answer or answer "Don't Know". Thus, if incorrect answers correspond to situations in which individuals believe they know the answer when in fact they don't, this can represent a measure of overconfidence. As the proportion of incorrect answers corresponds to a deviation from the "ideal" number of zero, it may be interpreted as a calibration error and used to estimate how confidence varies with knowledge. Figure 1B and Supplementary Figure S1 show some examples: 1) In the case of perfect metacognition, there should be no incorrect answers and no calibration error (yellow line); 2) The DKE would predict the fraction of incorrect answers to be higher in lower knowledge groups - dashed green lines); 3) If confidence did not vary with knowledge, the ratio should be similar for all knowledge levels, and in the particular case of random answering, both incorrect and "Don't Know" answers would have the same probability of occurring (dotted blue line); and 4) if over-confidence grew with knowledge, the proportion of incorrect answers should increase as well (purple line), although this has never been predicted in the literature. These scenarios give rise to monotonous relationships between confidence and knowledge and, except for the fourth, assume over-confidence to be independent from or decrease with knowledge. Therefore, all deviations from these predictions can be informative.

To eliminate the need for grouping respondents in quartiles, we applied our metric to three large-scale surveys on public understanding of science. These surveys (that we refer to as the EB [26], GSS [27], and Pew [28] datasets) asked several general science questions in the True/False/Don't Know format (or similar) to a total of 96039 respondents across the USA and 34 European territories, spanning a 30-year period (see Methods). For all three surveys, we first grouped respondents according to how many questions they answered correctly (yellow and x-axis in all figures) and secondly, for each knowledge bin (corresponding to the number of correctly answered questions), we grouped participants according to how their non-correct answers were distributed (all incorrect, all "Don't Know" or a combination of both). As our indirect confidence metric does not rely on comparison with peers (respondents are not asked to compare their performance to others on either axis), the results can be plotted in absolute scales (details in the Methods section). For each knowledge level, Figures 2A, C and E show the average proportion of correct (yellow), incorrect (purple) or "Don't Know" (green) answers across participants and Figures 2B, D and F show the distribution of participants with different proportions of "Don't Know" (green) and incorrect answers (purple). Respondents in all three surveys offer a sizable number of incorrect answers and these are not randomly distributed: the fraction of incorrect answers varies with knowledge: as the number of individuals who almost never offer wrong answers shrinks very fast this generates a nonlinear relationship, with respondents on both extremes (very low and very high knowledge) offering proportionally more "Don't Know" answers and being more likely to never answer incorrectly (Figure 2B, D and F dashed green lines and dark green bars). Following from our argument that the proportion of incorrect answers can be interpreted as a calibration error, both dashed lines in Figure 2 could be considered metrics of overconfidence: the green lines identify the fraction of individuals who rarely answered incorrectly (offered "Don't Know" answers in at least 80% of the cases), and the black lines the fraction that offered a similar number of incorrect and "Don't Know" answers, per knowledge bin (see Methods). As the black line offers a more conservative approach, we use it as a reference in this article and refer to it as the non-self-reported calibration error. We find that this calibration error is consistently lower for the least knowledgeable, in the three surveys, despite these surveys having covered different populations and spanning three decades, from the first EB to the Pew survey. This conflicts with both the perfect metacognition model, which would predict constant, zero calibration error, and with the DKE, which describes a higher confidence gap (or higher calibration error) at lower knowledge levels.

**Figure 2**

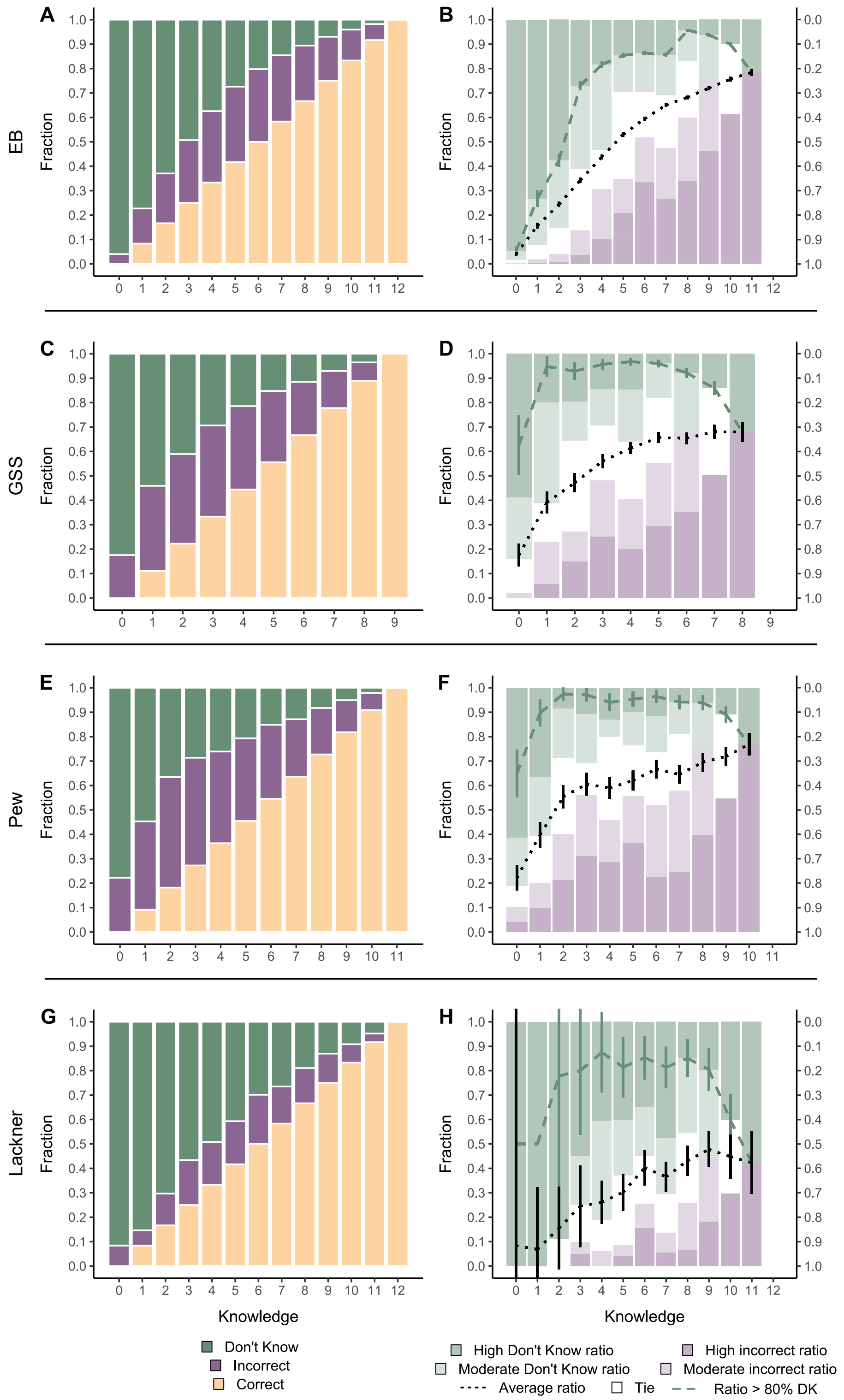

**Fig. 2. Overconfidence is higher at intermediate knowledge levels**. Data from EB (**A,B**), GSS (**C,D**), Pew (**E,F**) and Lackner (**G,H**) surveys. (**A,C,E,G**) The subgroups of each column show the average fraction of respondents answering "Don't know" (green), incorrectly (purple) or correctly (yellow), per knowledge level (number of questions answered correctly). (**B,D,F,H**) Each column shows the fractions of respondents according to knowledge level (proportion of "Don't Know" to incorrect answers by quintiles of normalized ratios): Dark green represents ≥ 0.8 normalized ratio (respondents with a majority of "Don't Know" answers), light green ≥ 0.6 to < 0.8, white ≥ 0.4 to < 0.6, light purple ≥ 0.2 to < 0.4 and dark purple 0 to < 0.2 (respondents with a majority of incorrect answers). Dotted black line shows

average normalized ratio and dashed green lines (read on right axis as the curve approximating the top of the graph represents a smaller fraction of “Don’t Know” answers) show fraction of participants with normalized ratios with a proportion of “Don’t Know” answers below 100% (the fraction of participants who had at least one incorrect answer). Bars correspond to 99% confidence intervals.

### The non-linear relation is independent of metric and robust across demographics

The observed non-linear relationship could be specific to our indirect metric, to non-controversial science-related questions, and/or to the analysed demographics. To test this, we followed four different approaches by: 1) comparing between different demographics; 2) developing a new survey with both our indirect confidence metric and a more traditional direct measure, 3) applying our metric to a previously and independently published study that reported evidence of the Dunning-Kruger effect in a controversial science-related topic [9], and 4) by simulating different answering strategies and testing the robustness of our metric to different knowledge distributions.

To compare between populations, we took advantage of the large span of the EB dataset and repeated the analysis per country, age, gender, and educational levels. We found very similar trends across the 34 territories and, as [29] before us, we identified the male middled-aged population as being more likely to never answer “Don’t Know” (Supplementary Materials, Supplementary Figure S2). However, as the EB does not offer stratified samples, the non-linear relationship could still be due to the studied demographics. Thus, we developed a new survey (referred to as “Lackner” throughout the paper). This survey focused on three countries (Portugal, Germany, and Norway), with different average education levels, and that displayed different percentages of “Don’t Know” avoidance (red in Supplementary Figure S2D) and compared different confidence metrics. We started by selecting a sample of respondents equally stratified in terms of age, gender, and education, and repeated the EB knowledge questionnaire with the True/False/Don’t Know structure. Figure 2G shows the proportion of correct (yellow), incorrect (purple) and “Don’t Know” (green) answers per knowledge level, and Figure 2H shows the distribution of incorrect and “Don’t Know” answers. As before, the proportion of incorrect answers is smaller at low knowledge levels (Figure 2H and 3E). We also included a direct confidence metric, resembling the format used in several DKE papers, by asking respondents to self-report the number of items they thought they answered correctly. Figure 3G shows these self-reported answers and Figure 3H the average confidence (dotted line). We also calculated the calibration error as the difference between the number of self-reported correct answers and the number of actually correctly-answered questions (Supplementary Figure S3). Once again, it can be observed that confidence grows very fast (and non-linearly) in the early knowledge bins for both confidence metrics (Figure 3F and 3H), all calibration error measures (Supplementary Figure S7) and in all countries analysed (Supplementary Figure S4). Interestingly, the trend in age, education and gender remained, indicating that some demographics might be particularly likely to overestimate their knowledge (Supplementary Figure S5).

As our study focuses on scientific knowledge (and we do not have access to the original DKE data), we used available data from a 2019 paper that included a True/False/Don’t Know knowledge questionnaire (referred to here as the Fernbach survey [9]), used a self-assessed, non-comparative confidence metric, and reported evidence of the DKE on the controversial topic of genetically modified foods. Again, when we plotted the proportion of incorrect (Figure 3J), we observed that this proportion grows very fast in the lower knowledge bins so that the least knowledgeable remain the more likely to answer “Don’t know” instead of incorrectly.

## Figure 3

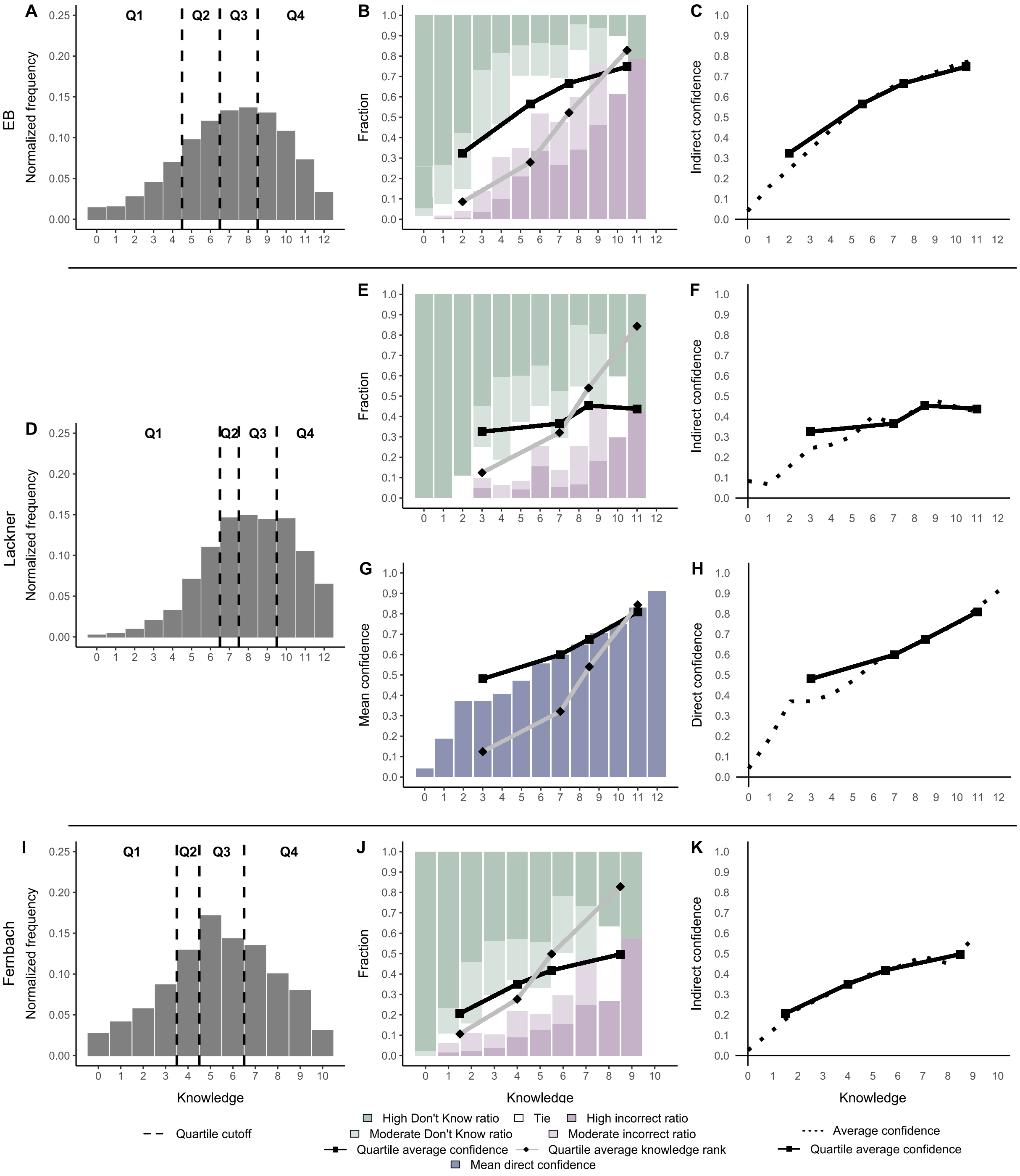

**Fig. 3. Non-linear relationship between confidence and knowledge is masked by binning.** Data from EB (A,B,C), Lackner (D,E,F,G,H), and Fernbach (I,J,K). **(A,D,I)** Frequency distributions of knowledge for the three datasets, with dashed vertical lines marking the quartiles. **(B,E,J)** Average fraction of non-correct answers that were "Don't Know" (green) or incorrect (purple) per knowledge level. **(G)** Average confidence (percentage of items self-reported as correct) per knowledge level. **(C,F,H,K)** Average confidence (dotted black line), calculated as the proportion of incorrect answers (C,F,K – equivalent to the dotted black lines in figure 2) or average percentage of items self-reported as correct (H). Grey lines with black diamonds in (B,E,G,J) show average knowledge ranking per knowledge quartile and black lines with black squares show average indirect confidence per quartile as average normalized ratios of incorrect to "Don't Know" answers in (B,E,J) and average percentage of items self-reported as correct in (G). Black lines with black squares in (C,F,H,K) show average indirect (C,F,K) and direct (H) confidence, per knowledge quartile.

Finally, we note that all described models (meta-cognition, DKE) assume not only that different metrics perfectly gauge knowledge but also that the metrics of confidence and knowledge are independent and this is rarely the case (for example, asking respondents to subjectively assess their knowledge before or after they have answered objective knowledge questions might alter their answers [52]). This lack of metric independence is particularly obvious for our indirect confidence metric as we are using the same questionnaire to gauge both knowledge and confidence: some respondents can correctly guess the answer when they in fact don't know, appearing more knowledgeable, and this might introduce a bias in our results. For example, when there are only three options (as in the EB survey, True, False, Don't Know) respondents who never guess and always answer "Don't Know" when they are unsure, would appear as less knowledgeable and simultaneously less confident, giving rise to the observed curves. On the other hand, if the number of options in the questionnaire increases (for example, four possible answers and one "Don't Know option, as in the Pew survey), guessing correctly becomes less likely and this effect should be less pronounced or null. Therefore, and to test whether this "guessing" effect could explain our results, we simulated different answering strategies, varying the average knowledge and the proportion of guessers in the population (Supplementary Figure S8). We found the parameters that best fit the knowledge distributions observed in all our surveys individually (EB, GSS, Pew, Lackner and Fernbach - Supplementary Figure S9) and compared the simulated proportions of incorrect answers to the observed proportion (Supplementary Figure S10) (details in the Methods). The best fits were obtained for approximately 25% of guessers (meaning that 25% of the simulated respondents always guess when they don't know) and, consistently, the respondents identified as the least knowledgeable are not the most overconfident, even when we correct for the presence of guessers (Supplementary Figure S11). In fact, the calibration error, measured here as the difference between the simulated (null model) and the observed ratio of incorrect answers, per knowledge bin, is very non-linear and peaks for intermediate knowledge levels. In our simulations, we can only reproduce a scenario in which the more over-confident are the least knowledgeable when we simultaneously decrease the mean knowledge of the population to very low values and increase the proportion of guessers to 75%, but we also lose the quality of the fit (Supplementary Figure S9), especially in the extreme bins. As anticipated above, this reversion is first observed in the Pew survey, that also has a very non-normal distribution of knowledge levels, something that we cannot explain (Supplementary Figures S5 and S9).

Therefore, and despite the limitations of the metric, we find that the effect is robust across countries, survey formats, scientific topics, distributions of knowledge, and proportions of guessers and that different answering strategies alone cannot explain the observed effect of lower confidence in the lower knowledge bins.

**A quadratic relationship is hidden by grouping the lower knowledge levels**

To our knowledge, this non-linear relationship between knowledge and confidence has not been described and cannot be predicted by the perfect metacognition model nor the Dunning-Kruger effect. It is robust and does not require comparative scales. As binning might be a relevant problem, particularly for skewed distributions and for extreme knowledge bins, we hypothesized that the DKE could be hiding important variability, by often relying on quartiles. Again, we tested this in two different ways. First, we plotted the knowledge distributions of all surveys (Figures 3A, D, I and Supplementary Figure S5) and confirmed that these are skewed to the right. Second, we repeated the analysis using our confidence metric, but representing knowledge and confidence in quartiles (Figures 3B, E and J, grey and black lines, respectively). Indeed, by using quartiles, the lower knowledge bins (five bins for the EB, seven for Lackner, and four for Fernbach) group together in the first quartile, and we now reproduce the DKE, with confidence varying little or growing approximately linearly between all knowledge levels, for both the indirect (Figure 3B, E and J) and direct confidence metrics (Figure 3G).

In summary, in all studies and for both confidence metrics, representing confidence using the quartile aggregation hides the observation that lower errors are found for low knowledge: the quadratic relation is revealed only when the analysis is done over the full knowledge levels (solid black line versus dotted black lines in Figure 3C, F, H, and K). It is important to note that this representation of the calibration error, using perfect-metacognition (or no incorrect answers) as the null-mode, penalizes the higher knowledge bins, with individuals who answered correctly all but one question appearing as over-confident. In fact, if we were to normalize by knowledge or use either random answering or linear growth of incorrect answers as null-models (Supplementary Figure S6), the biggest calibration errors would be found for intermediate knowledge bins only. Indeed, individuals in the most knowledgeable bins are more likely to offer almost exclusively "Don't know" answers than individuals in the intermediate knowledge bins, for all surveys (Figure 2B, D, F and H, green dashed line).

Altogether, this strongly suggests that, at least in the case of scientific knowledge, possessing some knowledge is more dangerous than having no knowledge, in the sense that the least knowledgeable are in fact aware of it and it is the ones with some knowledge who have less accuracy.

**Confidence modulates attitudes towards science**

One possible important consequence of over-estimating scientific knowledge pertains to public attitudes towards science, as recent studies on controversial science-related topics showed a role of overconfidence on negative attitudes towards science [8,9]. It is well-established that there is no linear relationship between knowledge and attitudes (unlike what the Deficit Model predicted [30,31]), and that such "attitudes" can vary widely depending on the subject, context, political identity, etc. [32-37] Therefore, and since we found the largest confidence gap at intermediate knowledge, we asked whether the least positive attitudes towards science would also be found in those knowledge levels.

Beyond the knowledge questionnaire, the EB also included a series of 10 independent attitudinal items, with answers that can be grouped in a Agree/Disagree/Don't Know format (see Methods). We first plotted the fraction of "Don't Know" answers per knowledge level and found that, for every attitude item, with only small variations, the least knowledgeable are the most likely to offer no opinion to the attitudinal questions (Supplementary Figure S8). Next, we analysed the attitude dependence on knowledge, and found that all relationships are quadratic or asymptotic (the effect was particularly strong in the "Agree" answers, possibly due to the acquiescence bias; see methods). This non-linear behaviour appears in all items, with what can be argued to be the most negative attitudes appearing at intermediate levels of knowledge (Figure 4A, C, E and Supplementary Figure S12), which also correspond to the highest confidence-to-knowledge ratios (shaded areas in Supplementary Figure S13).

Therefore, attitudes are neither independent from knowledge, nor do they appear to be more negative in lower knowledge levels, as the Deficit Model would predict. From our analysis, many of the attitudes that can be identified as negative seem to be modulated by a combination of some knowledge and confidence, with overconfidence appearing at the intermediate knowledge levels and reflecting itself in more negative attitudes towards science both in controversial and less-controversial topics. This has important implications for science communication and public compliance, discussed below.

**Figure 4**

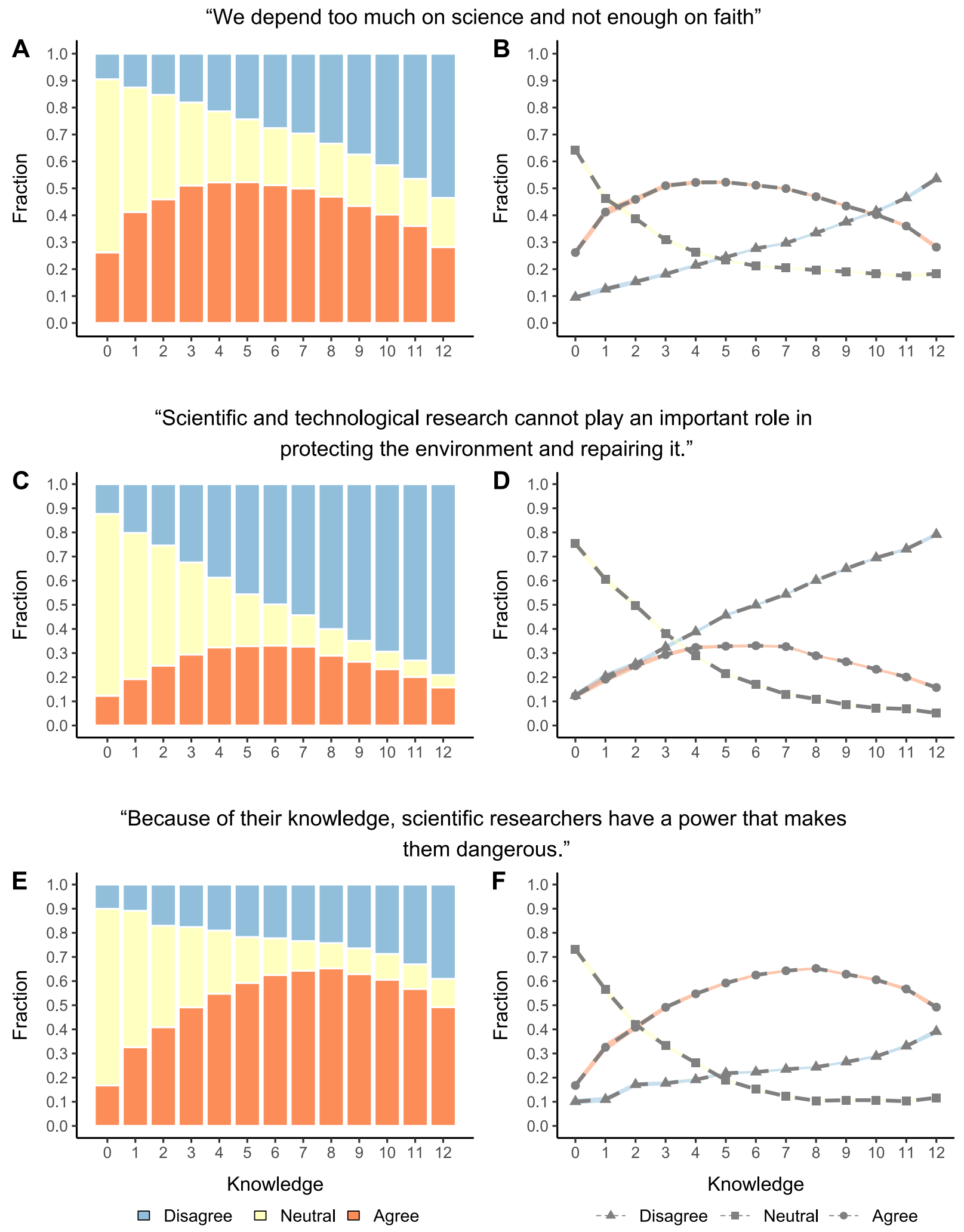

**Fig. 4. Negative attitudes toward science peak at intermediate knowledge levels.** EB data. Stacked bar plots with fractions of Agree (orange), Neutral (yellow) and Disagree (blue) answers for three different attitudinal questions: "We depend too much on science and not enough on faith" (A), "Scientific and technological research cannot play an important role in protecting the environment and repairing it" (C), "Because of their knowledge, scientific researchers have a power that makes them dangerous" (E). (B,D,F) show the mean fractions of each answer with shaded standard error of the mean for each respective question in red (Agree), yellow (Neutral) and blue (Disagree).

## DISCUSSION

In this paper we propose a new indirect, non-self-reported confidence metric, which does not rely on group comparison, and a new methodological approach, which looks at the full knowledge and confidence distributions. Using neutral answers as a proxy for confidence, both for the knowledge and the attitudinal questions, we found that 1) confidence grows much faster than knowledge, but 2) this growth is non-linear, with the largest confidence gaps at intermediate to high knowledge levels; 3) this effect is robust across metrics and countries; and 4) the least positive attitudes towards science are found for these high-confidence / intermediate-knowledge groups, in both controversial and non-controversial topics.

This is contrary to what was previously observed and, although not the first demonstration that a little knowledge can be a dangerous thing (see, e.g., [38]), might be due to different reasons, including methodological issues, namely that DKE-type studies often rely on quartile or model representations, and that previous analysis of EB attitudinal items mostly ignore the so-called neutral answers. In fact, more recent work by Sanchez and Dunning, provided evidence of a "beginners' effect", with confidence being low-to-moderate in the initial stages of learning a new task and growing very quickly after a few rounds of completing it, regardless of actual skill [39]. Despite being a different problem, as there is no correlation between number of trials and scientific knowledge, the fast increase in confidence and slower growth in skill or knowledge also results in non-linear relationships. Indeed, this effect might be less obvious in questionnaires where there is no rule to be learned and applied to other similar questions (such as deductive-reasoning tasks). Another possibility is that our method uses confidence and knowledge variables that have a similar number of items, whereas DKE studies often have more options on the knowledge/aptitude than on the confidence axis (e.g., 10 different grammar questions vs. one single "How do you compare with our peers on grammar"). Also, whether this effect is particular to scientific or another academic knowledge, remains to be tested. Finally, our indirect confidence metric uses the same questionnaire to gauge both confidence and knowledge and this raises important issues regarding metric independence. Our simulations suggest that this alone cannot explain the results but might introduce some bias. Taken as a whole, this study also highlights how all metrics have issues and using different null models and analytical strategies, including binning, might strongly impact the results. It adds to the important discussion of how to measure traits as complex as knowledge or confidence and the value of combining different tools, understanding the limitations of each.

Regarding the variation in attitudes towards science, we observed that the least knowledgeable were also more likely to offer "neutral" answers to the attitudinal questions, indicating lower confidence (Supplementary Figure S8). Previous studies [8,9] described the least positive attitudes in the lowest knowledge bins, but this difference might be due to the mentioned methodological differences in data binning and analysis. Still, this effect is much stronger in the EB dataset than in [8,9] and there can be three non-mutually exclusive explanations: 1) the EB surveyed many more people, unveiling important differences in the lower knowledge bins; 2) there is a significant time gap between these surveys and the EB dataset, that preceded the wide expansion of the internet and of online social networks. Misinformation and polarization have increased [40], possibly limiting the quality and diversity of accessible information, effectively creating large groups of misinformed citizens; 3) contrary to [8,9], the EB mostly focuses on non-controversial science issues, while respondents on more contested subjects, such as vaccination, might have access to more information on those subjects (both true and false), have stronger opinions, and believe themselves to be right (knowing the scientific consensus despite choosing not to follow it [41]). Moreover, the politicization of science, made obvious during the COVID-19 pandemic, together with an increase in political polarization [42,43] might deepen this divide. Therefore, if the EB described survey was to be repeated, we might observe differences in the gaps between confidence and knowledge, and possibly a stronger polarization in the attitude items. Thus, and despite the known problems of knowledge surveys [44], a new round of a very similar questionnaire should be considered.

Our results indicate that the least knowledgeable show at least some evidence of good metacognition, but that individuals with some knowledge are the most likely to overestimate it and to have less-positive attitudes towards science. Importantly, in all studies, most surveyed individuals fall on this intermediate knowledge and high confidence level (Figures 3 and Supplementary Figure S6): this effect was not important in our analysis, as bins were normalized by frequency, but is fundamental at a population level, as those intermediate groups are likely to correspond to a large demographic. Conversely, studying the extremes of the distribution was very difficult using classic surveys (with typically low numbers of respondents) but is increasingly possible in the social media and digital era, allowing for a deeper understanding of diversity, and identifying sub-populations.

This has clear implications for current science communication strategies. First, debiasing techniques should differ as a function of which model is correct: if the DKE holds, then interventions should target the most unaware, similar to the Deficit Model; if the present model holds, then they should target those with some intermediate knowledge (corresponding to the majority of the population). Second, our model indicates that receptiveness to science will be stronger at the lowest and highest knowledge levels, where the confidence-to-knowledge ratios are also lowest. Therefore, offering information that is incomplete, partial, or oversimplified, as science communicators often do, might backfire, as it may offer a false sense of knowledge to the public, leading to overconfidence, and less support, further reinforcing the negative cycle. Third, if the lowest support for science comes from the over-confident, these might also be the ones more resistant to new information, especially if it contradicts their certainty, creating a negative reinforcement loop. This resistance can manifest as confirmatory tendencies [45] or other cognitive biases. Thus, it is important to share accurate information, while also conveying humility, both on the scientists' and the lay public's side, and without colliding with individuals' values and ideology [46-49].

Together, our work suggests that, at least in the case of scientific topics, some knowledge is more dangerous than little knowledge, and it is fundamental to develop multidisciplinary approaches, building from psychology, social media, and complex systems analysis, to avoid such dangers.

## MATERIALS AND METHODS

### Experimental Design

This study takes advantage of large surveys in public understanding of science (existing or developed by the authors) to study how confidence varies with scientific knowledge. It introduces a new indirect confidence metric and tests it across many countries and years. All computations were performed using R 4.2.1, Microsoft Excel 16, Wolfram Mathematica 10 and Jupyter Notebook 6.01.

### Survey Datasets

Five different datasets were used, covering a large temporal range in Europe and the USA (see Supplementary Table TS1). The first three are large-scale surveys conducted by widely recognized entities, that focus on scientific knowledge and attitudes towards science and include scientific knowledge items in a True/False/Don't Know format or similar. The fourth survey was conducted by us, in 2021, in Germany, Portugal, and Norway. The fifth dataset is from a 2019 study on the Dunning-Krueger effect in a controversial science-related topic. For simplicity, we refer to these respectively as the EB, GSS, Pew, Lackner, and Fernbach throughout the text. The detailed items used for analysis are in Supplementary Tables TS2 (knowledge items) and TS3 (attitudinal items).

***Eurobarometer (EB):*** The EB dataset was obtained through five rounds of the Eurobarometer Science and Technology campaigns, from 1989 to 2005, surveying 34 territories, including EU members, candidates at the time, and other European Economic Area (EEA) countries, totalling 84469 individual interviews [26]. Unlike previous and subsequent campaigns, this set tried to gauge both knowledge and attitudes, in a consistent way. As there were differences both in the questions asked and in the possible answers, our dataset results from a harmonization effort that took the November 1992 (EB 38.1) round as a base and identified similar variables in the remaining four rounds [26]. This harmonized dataset is referred to as the EB dataset throughout the text.

***General Social Survey (GSS):*** The GSS dataset was obtained through seven rounds, bi-yearly from 2006 to 2018, surveying a panel of adults living in households in the United States (both English- and Spanish-language survey-takers) [27]. Datasets were homogenized based on knowledge question items, resulting in 7106 computer-assisted personal interviews. This dataset is referred to as the GSS dataset throughout the text.

***Pew Research Center's American Trends Panel (Pew):*** The Pew dataset was obtained through the Wave 42, conducted by Ipsos Public Affairs ("Ipsos") from January 7 to January 21, 2019, surveying a probability-based online panel of adults living in households in the United States, totalling 4464 online interviews [28]. This dataset is referred to as the Pew dataset throughout the text.

***This paper's study (Lackner):*** The Lackner dataset was obtained between April and May 2021 using Respondi (https://www.respondi.com) to recruit a stratified sample of respondents according to gender, age, and years of education or age at education completed, covering Portugal, Germany, and Norway. The survey was conducted online using the Qualtrics software. We received 1436 respondents total from which 442 respondents failed the data quality checks (Supplementary Table TS5), resulting in 994 respondents total (368 Portugal, 282 Norway, 344 Germany; details in Supplementary Methods). This dataset is referred to as the Lackner dataset throughout the text.

***Fernbach et al., study 2 (Fernbach):*** The Fernbach dataset was obtained from [9] through OSF (https://osf.io/t82j3/). The data collection took place in July 2016 using Qualtrics' panel and reached a sample of 1559 participants from France, Germany, and USA. While the original paper includes other studies, Study 2 had the largest sample and all knowledge questions referred to a single scientific knowledge area (genetics) which is relevant to a polarized topic (genetically modified foods). This dataset is referred to as the Fernbach dataset throughout the text.

**Knowledge Variables**

The different datasets have different knowledge items listed in Supplementary Table S2. The EB dataset includes 12 questions on general text-book science knowledge in a True / False / Don't Know format with the indication "If you don't know, say so, and we will skip to the next". While the EB dataset includes a 13rd item about how long the Earth takes to orbit the Sun, we chose to discard it as it was dependent on answering a previous question correctly. The GSS includes 9 questions in the exact same wording and True / False / Don't Know format as used for EB with the indication "If you don't know or aren't sure, just tell me so, and we will skip to the next question. Remember True, False, or Don't Know". The Pew includes 11 questions to test knowledge of science facts focusing on life science, earth and other environmental science as well as on applications of scientific principles, such as numeracy and chart reading, and the understanding of scientific processes. Answers are collected, differently to the EB and GSS, in a four-option multiple choice format and with the indication "If you don't know the answer, select Not sure". The Lackner survey included the same 12 questions of the EB and GSS in a True / False / Don't Know format with the indication "If you "Don't know", please say so". Finally, the Fernbach dataset includes 10 questions on genetics with a True / False / Don't Know format that were introduced with the instruction: "For each of the following statements, please tell us whether you think it is True or False".

We tested the knowledge items' independence by calculating Spearman correlations and by performing a Principal Components Analysis (PCA) using our largest database, EB (Supplementary Figures S14 and S15 and Supplementary Methods). As for the purposes of this project we were not so much interested in measuring individual knowledge as in finding relations between this measure and the other variables, we created a single knowledge variable, ranging from 0 (no correct answers) to 12 (all questions answered correctly). In the case of zero correct answers, these can have been answered incorrectly or as "Don't Know".

To obtain the knowledge quartiles, we calculated the three values that would result in an approximately equal division of the sample into four groups. As the number of items was reduced (12 in EB to 9 in GSS) and many participants had the same performance, the actual percentage in each quartile was not exactly 25%. As there was no reporting in the literature whether group intervals were typically closed on the left or on the right (e.g., are those participants who have a performance equal to Q1 excluded or included in the first group?), we decided to use intervals closed on the left consistently in all databases, as that guaranteed no oversampling of the lowest quartile (observed percentages ranged from 17% to 25%). For Figure 3, the knowledge quartile y-axis positioning (to be compared with the quartile's average confidence as in the Dunning-Kruger Effect studies) was calculated through the observed average knowledge ranking of participants in each quartile instead of assuming the theoretical quartile centroid (i.e., 12.5%, 37.5%, 62.5%, 87.5%).

**Confidence In Knowledge**

Confidence in one's knowledge can be measured in different ways but typically involves asking subjects directly (self-reporting). Common measures include asking subjects how knowledgeable they believe they are regarding specific issues, how well they believe they performed on a certain test (before seeing the actual test results) or asking them to compare (rate or rank) themselves to putative others performing the same test or task. Neither the EB, the Pew or the GSS included direct confidence metrics. We propose, and use in the paper, an indirect measure of confidence in knowledge, defined as the ratio of Incorrect to "Don't Know" answers in any knowledge questionnaire, as long as it had the format True / False / Don't Know (or similar). The rationale is that an incorrect answer corresponds to an overestimation of one's knowledge (more details in the main text). To normalize this ratio, we calculated the proportion of incorrect answers over all non-correct answers, as this allowed the data to be displayed in a way such that 0 would represent a ratio with no incorrect (e.g., 6 incorrect and 0 "Don't knows"), 1 would represent a ratio with only incorrect (e.g., 0 incorrect and 6 "Don't knows") and 0.5 would represent a tied ratio (e.g., 3 incorrect and 3 "Don't knows"), with meaningful in-between numbers (e.g., 0.33 would represent a proportion of 1:2 and all equivalent ratios such as 2:4 or 4:8).

To experimentally test the new indirect confidence measure, the Lackner survey included both this indirect metric and Overestimation, which is historically used to measure confidence in psychology and cognitive science. Respondents had to self-report how many of the total items they thought they got correctly ("Of the 12 questions you just answered, how many do you think you answered correctly?").

**Calibration Error Models**

To obtain an estimate of overconfidence, different proxies for calibration errors were defined. In the paper we assume a conservative approach and use as the null-model the perfect metacognition expectation: for the indirect confidence measure, calibration error is the difference between the expected proportion of "Don't Know" answers for a given knowledge level (e.g., when participants get no answers correct, all answers should be of the "Don't Know" type) and the actual observed proportion of "Don't Knows"- this corresponds to the mean proportion of Incorrect answers, per knowledge bin. Thus, calibration error varies between maximum overconfidence at 1 (e.g., if all answers were incorrect, we expect 100% "Don't Know" answers, so if the participant provided 0 "Don't Knows" = 1 - 0 = 1) and perfect calibration at 0 (the number of "Don't Knows" answers matches the expectation, e.g., when no correct answer was provided, we expect 100% "Don't Knows", 1 - 1 = 0). This is a very conservative metric as it does not correct for increasing knowledge (e.g., a person with a calibration error of 1 with 11 correct answers was overconfident in only 1 item, while a person with a calibration error of 1 with 0 correct answers was overconfident in 12 items). We discuss other possible null-models in the Supplementary Materials and Supplementary Figure S7 shows that in all of the cases, the lowest errors are found for the lowest knowledge bins. Please note that this indirect metric does not allow for explicit underconfidence (it is impossible to tell whether respondents who answered "Don't Know" actually knew the answer), so the metric might underestimate overconfidence.

**Simulations**

To gauge the impact of guessing on the knowledge distribution (and whether this could explain the observed low confidence at low knowledge levels) different answering strategies were simulated using R 4.2.1, using ggplot2 for the figures. This was done in five steps.

First, we generated a baseline knowledge distribution to represent the "true knowledge distribution" of the agents. EB, GSS and Lackner displayed a bell-like distribution and Pew seemed to follow a more linear pattern. Following form this observation we opted to simulate an approximate Normal and approximate one-sided Triangular distributions. For the Normal distribution, this was done by randomly taking 100,000 points, with mean (*M*) and standard deviation (*SD*) as variable parameters. The parameter space was explored by varying both *M* and *SD* and starting from the observed parameters in each survey (EB, GSS, Pew, Lackner and Fernbach). As guessing inflates the knowledge distribution, there is little point in exploring higher "true" knowledge scenarios, so parameter estimation of *M* was asymmetrical, testing mostly scenarios in which true average K was lower than observed. Therefore, *M* was varied in ten steps of 0.5 (1 above and 9 below), unless it reached zero, and *SD* was varied in 6 steps of 0.5, three above and three below the initial observed *SD*, unless it reached zero. In the case of the Pew dataset, which displayed an obviously non-quadratic distribution, a one-side, two parameter triangular distribution was also tested and we used a regression equation to calculate the frequency of each knowledge bin, with $b_0$ (intercept) and $b_1$ (slope) as variable parameters. As before, we started from the observed $b_0$ and $b_1$ and tested 10 steps for $b_0$ and 6 steps for $b_1$ of 0.005.

Second, we simulated the “observed knowledge distribution”, by adding agents who guess when they don’t know the answer and guess correctly with a frequency proportional to the number of options (2, or 50% in the case of the EB, GSS, Lackner and Fernbach and 4, or 25%, for Pew). These correct answers were added to that agent’s total knowledge score, who would appear to be more knowledgeable than the agents who never guessed. To vary the proportion of guessers in the population, we created “combined knowledge distributions” by weighing the sum of the two previous distributions. The weight was either 0.25, 0.5 or 0.75, corresponding to the percentage of agents who would always guess when did not know the correct answer. The frequency of agents who always answered DK was weighted by 1 – this parameter (Supplementary Figure S9).

Third, we estimated the proportion of DK to incorrect answers for each knowledge bin, by finding the proportion of agents that came directly from the DK distribution (of step 1) and the proportion of agents that came from the guessing distribution (of step 2), for the different weighting processes. Thus, the proportion of guessers at a given bin represents the proportion of agents in that bin who would have used a guessing strategy (so that all other answers they provided were incorrect answers) and the proportion of agents that always say DK at a given bin represents the proportion of agents in that bin who would have always said DK when faced with an unknown question (so that all other answers the provided are of the DK type).

Finally, the “combined knowledge distribution” was compared to the actual knowledge distribution observed in each dataset by calculating the MSE for each combination of parameters. A heatmap with the different MSE can be found in Supplementary Figure S8. The parameter and distribution type combination with lowest MSE for each dataset is presented in (Supplementary Figure S9).

**Attitude Variables and Confidence in Attitudes**

The EB dataset contains 10 core attitude variables listed in Supplementary Table S3. The number of attitude questions in each EB round varied from year to year but there was a core of 10 questions homogenized from all rounds that have been used in subsequent studies. Since the GSS survey only included 3 of the 10 attitude questions in the EB survey (marked with asterisk in Supplementary Table S3) and the Pew survey used different items, the 10 EB core questions were selected for analysis and repeated in the Lackner survey. The possible answers to the attitude questions were not consistent over time and were systematized (Supplementary Methods): in the main text, “Neutral Answers” represent the aggregates of “Neither Agree Nor Disagree” and “Don’t Know” answers. Similarly to what was done for the knowledge questions, the “Agree” and “Disagree” answers to the attitudinal questions were compared to the proportion of “Neutral” answers.

To identify a possible "attitude" dimension, we expanded the work of [34] and included all Eurobarometers and territories, offering more data and statistical power, and the possibility of comparing the results longitudinally. Spearman correlation showed poor correlation between items and the first 5 components of a PCA represented around 65% of the variances (Supplementary Methods and Supplementary Figures S14 and S15). Thus, all attitude variables were treated independently. Regardless of the polarity of the questions, "Agree" and "Strongly Agree" answers were typically more prevalent than disagreement answers (a common effect, known as "acquiescence bias" [33,50]); therefore, the Results section focuses on the "Agree" answers, as these tend to show a more obvious effect.

## REFERENCES

1. S. Plous, *The Psychology of Judgment and Decision Making* (McGraw-Hill, 1993).
2. S. Lichtenstein, B. Fischhoff, Do those who know more also know more about how much they know? *Organ. Behav. Hum. Perf.* **20**, 159-183 (1977).
3. D. D. P. Johnson, S. A. Levin, The tragedy of cognition: psychological biases and environmental inaction. *Curr. Sci*. **97**, 1593-1603 (2009).
4. B. W. Tuchman, *The March of Folly: From Troy to Vietnam* (Alfred A. Knopf, 1984).
5. D. D. P. Johnson, D. R. Tierney, The Rubicon theory of war: how the path to conflict reaches the point of no return. *Int. Secur.* **36**, 7-40 (2011).
6. G. A. Akerlof, R. J. Shiller, *Animal Spirits: How Human Psychology Drives the Economy, and Why it Matters for Global Capitalism* (Princeton Univ. Press, 2009).
7. J. Kruger, D. Dunning, Unskilled and unaware of it: how difficulties in recognizing one's own incompetence lead to inflated self-assessments. *J. Pers. Soc. Psychol.* **77**, 1121-1134 (1999).
8. M. Motta, T. Callaghan, S. Sylvester, Knowing less but presuming more: Dunning-Kruger effects and the endorsement of anti-vaccine policy attitudes. *Soc. Sci. Med.* **211**, 274-281 (2018).
9. P. M. Fernbach, N. Light, S. E. Scott, Y. Inbar, P. Rozin. Extreme opponents of genetically modified foods know the least but think they know the most. *Nat. Hum. Behav.* **3**, 251-256 (2019).
10. I. S. Al-Harthy, C. A. Was, A. S. Hassan, Poor performers are poor predictors of performance and they know it: can they improve their prediction accuracy. *Journal of Global Research in Education and Social Science* **4**, 93-100 (2015).
11. M. Händel, E. S. Fritzsche, Unskilled but subjectively aware: metacognitive monitoring ability and respective awareness in low-performing students. *Mem. Cognition* **44**, 229-241 (2016).
12. T. M. Miller, L. Geraci, Unskilled but aware: reinterpreting overconfidence in low-performing students. *J. Exp. Psychol. Learn.* **37**, 502-506 (2011).
13. M. Urban, K. Urban, Unskilled but aware of it? Cluster analysis of creative metacognition from preschool age to early adulthood. *J. Creative Behav.* **55**, 937-945 (2021).
14. C. Sanchez, D. Dunning, Cultural patterns explain the worldwide perception/performance paradox in student self-assessments of math and science skill. *Soc. Psychol. Pers. Sci.* **10**, 935-945 (2019).

15. R. D. McIntosh, E. A. Fowler, T. Lyu, S. Della Sala, Wise up: clarifying the role of metacognition in the Dunning-Kruger effect. *J. Exp. Psychol. Gen.* **148**, 1882-1897 (2019).

16. G. E. Gignac, M. Zajenkowski, The Dunning-Kruger effect is (mostly) a statistical artefact: valid approaches to testing the hypothesis with individual differences data. *Intelligence* **80**, 101449 (2020).

17. R. A. Jansen, A. N. Rafferty, T. L. Griffiths, A rational model of the Dunning-Kruger effect supports insensitivity to evidence in low performers. *Nat. Hum. Behav.* **5**, 756-763 (2021).

18. A. J. Nederhof, Methods of coping with social desirability bias: a review. *Eur. J. Soc. Psychol.* **15**, 263-280 (1985).

19. R. B. Larson, Controlling social desirability bias. *Int. J. Market Res.* **61**, 534-547 (2019).

20. J. S. Cole, R. M. Gonyea, Accuracy of self-reported SAT and ACT test scores: implications for research. *Res. High. Educ.* **51**, 305-319 (2010).

21. A. G. Greenwald, T. A. Poehlman, E. L. Uhlmann, M. R. Banaji, Understanding and using the Implicit Association Test: III. Meta-analysis of predictive validity. *J. Pers. Soc. Psychol.* **97**, 17-41 (2009).

22. G. F. Bishop, A. J. Tuchfarber, R. W. Oldendick, Opinions on fictitious issues: the pressure to answer survey questions. *Public Opin. Quart.* **50**, 240-250 (1986).

23. C. Sedikides, M. D. Alicke, “Self-enhancement and self-protection motives” in *The Oxford Handbook of Human Motivation*, R. M. Ryan, Ed. (Oxford Uni., 2012), chap. 17.

24. W. De Neys, Bias and conflict: a case for logical intuitions. *Perspect. Psychol. Sci.* **7**, 28-38 (2012).

25. E. F. Williams, D. Dunning, J. Kruger, The hobgoblin of consistency: algorithmic judgment strategies underlie inflated self-assessments of performance. *J. Pers. Soc. Psychol.* **104**, 976-994 (2013).

26. M. W. Bauer, R. Shukla, P. Kakkar, “Public understanding of science in Europe 1989-2005. A Eurobarometer trend file” (Tech. Rep. Gesis ZA4669, European Commission, 2012).

27. T. W. Smith, M. Davern, J. Freese, S. L. Morgan, “General social surveys, 1972-2018” (NORC, 2019).

28. “American trends panel wave 42”, Pew Research Center (2019).

29. M. Bauer, Socio-demographic correlates of DK-responses in knowledge surveys: self-attributed ignorance of science. *Soc. Sci. Inform.* **35**, 39-68 (1996).

30. J. R. Durant, G. A. Evans, G. P. Thomas, The public understanding of science. *Nature* **340**, 11-14 (1989).

31. M. W. Bauer, N. Allum, S. Miller, What can we learn from 25 years of PUS survey research? liberating and expanding the agenda. *Public Understand. Sci.* **16**, 79-95 (2007).

32. B. Wynne, Knowledges in context. *Science, Technology, & Human Values*, **16**, 111-121 (1991).

33. G. Evans, J. Durant, The relationship between knowledge and attitudes in the public understanding of science in Britain. *Public Understand*. *Sci.* **4**, 57-74 (1995).

34. R. Pardo, F. Calvo, Attitudes toward science among the European public: a methodological analysis. *Public Understand. Sci.* **11**, 155-195 (2002).

35. L. C. Hamilton, Education, politics and opinions about climate change evidence for interaction effects. *Climatic Change* **104**, 231-242 (2011).

36. A. M. McCright, Political orientation moderates Americans' beliefs and concern about climate change. *Climatic Change* **104**, 243-253 (2011).

37. C. Drummond, B. Fischhoff, Individuals with greater science literacy and education have more polarized beliefs on controversial science topics. *P. Natl. Acad. Sci. USA*, **114**, 9587-9592 (2017).

38. S. Oskamp, Overconfidence in case-study judgments. J. Couns. Psychol. 29, 261-265 (1965).

39. C. Sanchez, D. Dunning, Overconfidence among beginners: is a little learning a dangerous thing? *J. Pers. Soc. Psychol.* **114**, 10-28 (2018).

40. W. Quattrociocchi *et al.*, http://dx.doi.org/10.2139/ssrn.2795110 (2016).

41. National Science Board, "Science and technology: public attitudes and understanding", *Science and Engineering Indicators 2014* (National Science Foundation, 2014), chap. 7.

42. S. Iyengar, M. Krupenkin, The strengthening of partisan affect. *Polit. Psychol.* **39**, 201-218 (2018).

43. E. Klein, *Why We're Polarized* (Profile Books Ltd., 2020).

44. M. Bauer, "Survey research and the public understand of science" in *Handbook of Public Communication of Science and Technology*, M. Bucchi, B. Trench, Eds. (Routledge, 2008), pp. 111-129.

45. R. S. Nickerson, Confirmation bias: a ubiquitous phenomenon in many guises. *Rev. Gen. Psychol.,* **2**, 175-220 (1998).

46. T. H. Campbell, A. C. Kay, Solution aversion: on the relation between ideology and motivated disbelief. *J. Pers. Soc. Psychol.* **107**, 809-824 (2014).

47. M. Feinberg, R. Willer, The moral roots of environmental attitudes. *Psychol. Sci.* **24**, 56-62 (2013).

48. M. Feinberg, R. Willer, From gulf to bridge: when do moral arguments facilitate political influence? *Pers. Soc. Psychol. B.* **41**, 1665-1681 (2015).

49. M. J. Hornsey, Why facts are not enough: understanding and managing the motivated rejection of science. *Curr. Dir. Psychol. Sci.* **29**, 583-591 (2020).

50. R. Johns, Likert items and scales. *Survey question bank: Methods fact sheet* **1**, 11-28 (2010).

51. N. Light, P. M. Fernbach, N. Rabb, M. V. Geana, & S. A. Sloman, Knowledge overconfidence is associated with anti-consensus views on controversial scientific issues. *Science Advances* **8**, eabo0038 (2022).

52. R. H. Maki, "Test predictions over text material", *Metacognition in Educational Theory and Practice* (Routledge, 1998), chap. 6.

## Acknowledgments

The authors thank Caetano Souto-Mayor, Michael West and João Nolasco for initial extraction and analysis of the EB dataset, members of the SPAC group for valuable discussions, Fernbach et al. [9] for making their survey data available, and Martin Bauer, Tiago Paixão, Marta Entradas and Joana Lobo Antunes for critical reading of the manuscript. This project was partially funded by Welcome DFRH WIIA 60 2011 and ERC-Starting Grant FARE-853566, both to JGS.

**Ethical compliance**:

We have complied with all relevant ethical regulations and a Data Protection Impact Assessment was evaluated by a certified DPO. The Lackner survey obtained ethical clearance from the Nova SBE's Scientific Council (where the corresponding author was previously located and where the study started), following independent advice from its CICE- installation committee of the ethics review board, reference 13/2020, from 25/03/2020. The Lackner survey participants were paid through a third party (Respondi) after obtaining informed consent.

**Data Availability**

Surveys EB, Pew and GSS are publicly available and data and details can be found in [26], [27] and [28], respectively. The Fernbach study was published in [9] and the authors made the data available. Lackner survey data is available here: https://doi.org/10.5281/zenodo.7920776

Code Availability

All code used for the analysis is available here: https://doi.org/10.5281/zenodo.7920750

**Acknowledgments**

The authors thank Caetano Souto-Mayor, Michael West and João Nolasco for initial extraction and analysis of the EB dataset, members of the SPAC group for valuable discussions, Fernbach et al. [9] for making their survey data available, and Martin Bauer, Tiago Paixão, Marta Entradas and Joana Lobo Antunes for critical reading of the manuscript. This project was partially funded by Welcome DFRH WIIA 60 2011 and ERC-Starting Grant FARE-853566, both to JGS. The funders had no role in study design, data collection and analysis, decision to publish or preparation of the manuscript.

**Competing interests:** Authors declare that they have no competing interests.

**Supplementary Materials:** Annexed document

# SUPPLEMENTARY MATERIALS

# SUPPLEMENTARY METHODS

## OTHER DATASETS

***Latitudes:*** to get coordinates for each territory's centroids, built-in natural earth data by Geopandas in Python was used (for more information http://geopandas.org/). Malta was not included in Geopandas and was excluded for further analysis. West and East Germany were combined, as well as Northern Ireland and Ireland.

## KNOWLEDGE VARIABLE INDEPENDENCE

We tested the EB knowledge items' independence by calculating Spearman correlations and by performing a Principal Components Analysis (PCA) (Supplementary Figures S8B and S9B). The knowledge answers were correlated, albeit poorly, and this can be explained in great part by their different difficulty levels, with some questions displaying a much higher number of correct answers. Also, contrary to the attitude questions (details below), the knowledge items PCA revealed that the first component explains 25% of the variance, with all components having the same sign. This indicates that answering one question correctly increases the likelihood of giving the right answer to other questions (Figure S9B). In fact, the distribution of correct answers is approximately Normal, as expected, but skewed to the right (Figure 3A).

## ATTITUDES VARIABLES HARMONIZATION AND INDEPENDENCE

The EB dataset [1] contains 10 core attitude variables listed in Supplementary Table S3. The possible answers to the attitude questions were not consistent: 1) the "Don't Know" option was always present but a neutral option such as "Neither Agree Nor Disagree" was only offered in 1989, 1992 and 2005; 2) the available options on the Likert scale were sometimes five and others two, as shown in Supplementary Table S4. As these differences may have an impact on the respondents' behaviour [2], we tested this possible impact in three different ways: 1) by treating all the categories in the Likert scale either separately or fusing them into less options (adding the "Strongly Agree" with the "Agree to some extent", and the "Disagree to Some Extent" with the "Strongly Disagree"); 2) by either including or disregarding the "Neither Agree Nor Disagree"; and 3) by either aggregating the "Neither Agree nor Disagree" with the "Dont' Know" answers, or by treating them separately. These alternatives make up for a total of six different approaches and we performed many of the calculations in all six ways with no significant differences.

We then tested all 10 items for independence through Spearman correlation and Principal Components Analysis. We found that they are weakly correlated (< 0.33), with only two groups of variables with relatively higher correlations: one that might be associated with an optimistic attitude and another with overall distrust, as shown in Figure S8A. Second, we performed a PCA and found, as [2] before us, that this system does not justify the grouping of some attitudinal questions, as can be seen in Figure S9A. Indeed, the first and most significant principal component accounts for less than 20% of the variance and even the first 5 components only represent around 65%, with the last and less significant of 10 components still holding almost 5% of the variance. Therefore, all items were treated independently in the analysis.

## ROBUSTNESS TO DIFFERENT DEMOGRAPHICS

As the EB survey includes over 84000 respondents, we repeated the indirect confidence metric analysis in two different ways: first, we calculated the I/DK ratio for each of the 34 territories individually, and it peaks at intermediate knowledge levels in all surveyed territories; second, we focused on the large subset of respondents who never answered “Don’t Know” and found they were more likely to be male, middle-aged, have intermediate education, and become more common as latitude increases (Supplementary Figure S1 A, B, C and D). In the case of the 2021 Lackner survey, that used these similarly stratified samples for Portugal, Norway and Germany, we could no longer see a latitude effect (Supplementary Figure S2), indicating that the previously identified differences were mostly due to different education levels, and not necessarily to cultural differences. However, the trend in age, education and gender remained, indicating that some demographics might be particularly likely to overestimate their knowledge (Supplementary Figure S3).

## LACKNER 2021 SURVEY DATA QUALITY

Criteria for exclusion: Four different attention checks (reading comprehension, repeated knowledge item, self-reported attention, self-reported data quality) were included in the survey (Table S5). The following three criteria were used to drop participants: (1) We dropped all respondents that failed the first attention check in the first third of the survey that explicitly calls out for paying attention. (2) We dropped all respondents that failed two or more of the remaining three attention checks. (3) The survey was expected to take 15 min reading time (250 words per min). We dropped all respondents that were faster than half that time (7.5min).

## SUPPLEMENTARY REFERENCES

[1] Bauer, M. W., Shukla, R., & Kakkar, P. Public understanding of science in Europe 1989-2005. A Eurobarometer trend file. *Tech. Rep.* (2012).
[2] Pardo, R., & Calvo, F. Attitudes toward science among the European public: a methodological analysis. *Public Understand. Sci.* 11.2, 155-195 (2002).

# SUPPLEMENTARY TABLES

Table S1. List of territories, data collection dates and final sample size for the five databases used.

| **Survey** | **1) EB** | **2) GSS** | **3) Pew** | **4) Lackner** | **5) Fernbach** |
|---|---|---|---|---|---|
| **Dates** | 1989-2005 | 2006-2018 | Jan 2019 | Apr-May 2021 | Jul 2016 |
| **Final sample size** | 84469 | 7106 | 4464 | 994 | 1559 |
| **Austria** | 2 | - | - | - | - |
| **Belgium** | 4 | - | - | - | - |
| **Bulgaria** | 2 | - | - | - | - |
| **Croatia** | 1 | - | - | - | - |
| **Cyprus** | 2 | - | - | - | - |
| **Czech Republic** | 2 | - | - | - | - |
| **Denmark** | 4 | - | - | - | - |
| **Estonia** | 2 | - | - | - | - |
| **Finland** | 2 | - | - | - | - |
| **France** | 4 | - | - | - | 1 |
| **Germany – East** | 3 | - | - | 1* | 1* |
| **Germany – West** | 4 | - | - | 1* | 1* |
| **Great Britain** | 4 | - | - | - | - |
| **Greece** | 4 | - | - | - | - |
| **Hungary** | 2 | - | - | - | - |
| **Iceland** | 1 | - | - | - | - |
| **Ireland** | 4 | - | - | - | - |
| **Italy** | 4 | - | - | - | - |
| **Latvia** | 2 | - | - | - | - |
| **Lithuania** | 2 | - | - | - | - |
| **Luxembourg** | 4 | - | - | - | - |
| **Malta** | 2 | - | - | - | - |
| **Netherlands** | 4 | - | - | - | - |
| **Northern Ireland** | 4 | - | - | - | - |
| **Norway** | 1 | - | - | 1 | - |
| **Poland** | 2 | - | - | - | - |
| **Portugal** | 4 | - | - | 1 | - |
| **Romania** | 2 | - | - | - | - |
| **Slovakia** | 2 | - | - | - | - |
| **Slovenia** | 2 | - | - | - | - |
| **Spain** | 4 | - | - | - | - |
| **Sweden** | 2 | - | - | - | - |
| **Switzerland** | 1 | - | - | - | - |
| **Turkey** | 2 | - | - | - | - |
| **United States** | - | 7 | 1 | - | 1 |
| ***N* unique territories** | 34 | 1 | 1 | 3 | 3 |

West and East Germany are only treated as separate territories in the EB database.

* Note: Numbers indicate how many times a given territory was included in different data collection waves for a given database.

Table S2. List of science knowledge questions used for data analysis from each of the five surveys

Eurobarometer (1), General Social Survey (2), Pew (3), Lackner (4) and Fernbach (5).

| **Question** | **Answer options** | **Surveys** |
|---|---|---|
| “The centre of the Earth is very hot.” | **“True**”, “False” | 1, 2, 4 |
| “The oxygen we breathe comes from plants.” | “**True**”, “False” | 1, 4 |
| “Radioactive milk can be made safe by boiling it.” | “True”, “**False**” | 1, 4 |
| “Electrons are smaller than atoms.” | “**True**”, “False” | 1, 2, 4 |
| “The continents on which we live have been moving their location for millions of years and will continue to move in the future.” | “**True**”, “False” | 1, 2, 4 |
| “It is the father’s gene which decides whether the baby is a boy or a girl.” | “**True**”, “False” | 1, 2, 4* |
| “The earliest humans lived at the same time as the dinosaurs.” | “True”, “**False**” | 1, 4 |
| “Antibiotics kill viruses as well as bacteria.” | “True”, “**False**” | 1, 2, 4 |
| “Lasers work by focusing sound waves.” | “True”, “**False**” | 1, 2, 4 |
| “All radioactivity is man-made.” | “True”, “**False**” | 1, 2, 4 |
| “Human beings, as we know them today, developed from earlier species of animals.” | “**True**”, “False” | 1, 2, 4 |
| “Does the earth go around the sun or does the sun go around the earth?” | “The sun goes around the earth”, “**The earth goes around the sun**” | 1, 2, 4 |
| “Oil, natural gas and coal are examples of…“ | “Biofuels”, “**Fossil fuels**”, “Geothermal resources”, “Renewable resources” | 3 |
| “A scientist is conducting a study to determine how well a new medication treats ear infections. The scientist tells the participants to put 10 drops in their infected ear each day. After two weeks, all participants’ ear infections had healed. Which of the following changes to the design of this study would most improve the ability to test if the new medication effectively treats ear infections?” | “**Create a second group of participants with ear infections who do not use any ear drops**”, “Create a second group of participants with ear infections who use 15 drops a day”, “Have participants use ear drops for only 1 week”, “Have participants put ear drops in both their infected ear and healthy ear” | 3 |
| “Which of the following is an example of genetic engineering?“ | “Growing a whole plant from a single cell”, “Finding the sequences of bases in plant DNA”, “**Inserting a gene into plants that makes them resistant to insects**”, “Attaching the root of one type of plant to the stem of another type of plant“ | 3 |
| “What is the main cause of seasons on the Earth?” | “The distance between the Earth and the Sun”, “**The tilt of the Earth’s axis in relation to the Sun”,** “The speed that the Earth rotates around the Sun”, “Changes in the amount of energy coming from the Sun” | 3 |
| “These graphs show the monthly precipitation and average temperature for three cities in the United States over the course of one year. Based | “New York, New York”, “Los Angeles, California”, “**Chicago, Illinois**”, “They | 3 |
| on the graphs, which city has the greatest annual range of temperatures?” | all have the same annual temperature range” | |
| “The time a computer takes to start has increased dramatically. One possible explanation for this is that the computer is running out of memory. This explanation is a scientific…” | “Conclusion”, “Experiment”, “**Hypothesis**”, “Observation” | 3 |
| “Many diseases have an incubation period. Which of the following best describes what an incubation period is?” | “The recovery period after being sick”, “The effect of a disease on babies”, “The period during which someone builds up immunity to a disease”, “**The period during which someone has an infection, but is not showing symptoms**” | 3 |
| “When large areas of forest are removed so land can be converted for other uses, such as farming, which of the following occurs?” | “**Increased erosion**”, “Colder temperature”, “Decreased carbon dioxide”, “Greater oxygen production” | 3 |
| “An antacid relieves an overly acidic stomach because the main components of antacids are…“ | “Acids”, “Neutral”, “**Bases**”, “Isotopes” | 3 |
| “Which of these is a major concern about the overuse of antibiotics?” | “**It can lead to antibiotic-resistant bacteria**”, “There will be an antibiotic shortage”, “Antibiotics can cause secondary infections”, “Antibiotics will get into the water system” | 3 |
| “A car travels at a constant speed of 40 miles per hour. How far does the car travel in 45 minutes?” | “25 miles“, “**30 miles**”, “35 miles”, “40 miles” | 3 |
| “Yeast for brewing beer or making wine consists of living organisms.” | “**True**”, “False” | 5 |
| “Ordinary tomatoes do not contain genes, while genetically modified tomatoes do.” | “True”, “**False**” | 5 |
| “The cloning of living things produces genetically identical copies.” | “**True**”, “False” | 5 |
| “By eating a genetically modified fruit, a person's genes could also become modified.” | “True”, “**False**” | 5 |
| “It is possible to find out in the first few months of pregnancy whether a child will have Down's Syndrome.” | “**True**”, “False” | 5 |
| “Genetically modified animals are always bigger than ordinary ones.” | “True”, “**False**” | 5 |
| “More than half of human genes are identical to those of a chimpanzee.” | “**True**”, “False” | 5 |
| “It is not possible to transfer animal genes into plants.” | “True”, “**False**” | 5 |
| “Human cells and human genes function differently from those in animals and plants.” | “True”, “**False**” | 5 |
| “Embryonic stem cells have the potential to develop into normal humans.” | “**True**”, “False” | 5 |

* Note: From Eurobarometer 2005 onward the word “father” was replaced with “mother” in this question and the correct answer became “False”. The Lackner survey dataset also used the mother’s gene formulation.

Table S3. List of attitudes towards science questions used for data analysis from the Eurobarometer (EB) dataset.

For each statement respondents were asked to state their agreement or disagreement. Starred items (∗) indicate items present in General Social Survey. Items marked with a dagger (†) were not part of EB in 1989.

| **Question** |
| --- |
| *"Science & Technology are making our lives healthier, easier and more comfortable." |
| †"Thanks to scientific and technological advances, the earth's natural resources will be inexhaustible." |
| "We depend too much on science and not enough on faith." |
| †"Scientific and technological research cannot play an important role in protecting the environment and repairing it." |
| †"Scientists should be allowed to do research that causes pain and injury to animals like dogs and chimpanzees if it can produce information about human health problems." |
| †"Because of their knowledge, scientific researchers have a power that makes them dangerous." |
| †"The application of science and new technology will make work more interesting." |
| †"For me, in my daily life, it is not important to know about science." |
| *"Science makes our way of life change too fast." |
| †*"Thanks to science and technology, there will be more opportunities for the future generations." |

Table S4. Available answers for attitude items in each Eurobarometer (EB) campaign.

| | **EB 31 Mar-Apr 1989** | **EB 38.1 Nov 1992** | **EB 55.2 May-Jun 2001** | **Candidate EB 2002.3 Oct-Nov 2002** | **EB 63.1 Jan-Feb 2005** |
|---|---|---|---|---|---|
| **Strongly agree** | • | • | - | - | • |
| **Agree to some extent** | • | • | • | • | • |
| **Neither agree nor disagree** | • | • | - | - | • |
| **Disagree to some extent** | • | • | • | • | • |
| **Strongly disagree** | • | • | - | - | • |
| **Don't know** | • | • | • | • | • |

Table S5. Attention checks deployed in the Lackner survey.

| **Attention checks** | **Description** |
|---|---|
| 1 | "It is important that you pay attention to this study, please tick "Disagree" to show that you are reading the questions carefully."<br>Answer options: 5-point scale from "Strongly Disagree" to "Strongly Agree" with "Don't know" option. |
| 2 | The science knowledge item "Antibiotics kill viruses as well as bacteria" was presented twice. This attention check was considered failed if the answer was different in the two presentations.<br>Answer options: "True", "False", "Don't know" |
| 3 | "This is an attention check to understand the quality of your data.  Please be honest, you will get your reward regardless of your response. Did you respond randomly at any point during the study? Please be honest, you will get your reward regardless of your response."<br>Answer options: "Yes", "No" |
| 4 | "Is there any reason why we should consider your data with caution? (For example: you were listening to music during the study, you took a break to go to the bathroom, you were watching TV or were on Facebook at the same time...). Please be honest, you will get your reward regardless of your response."<br>Answer options: "Attention: You should consider my data with caution.", "You may consider my data: I completed the study without interruptions." |

# SUPPLEMENTARY FIGURES

## Supplementary Figure 1. Model Comparison

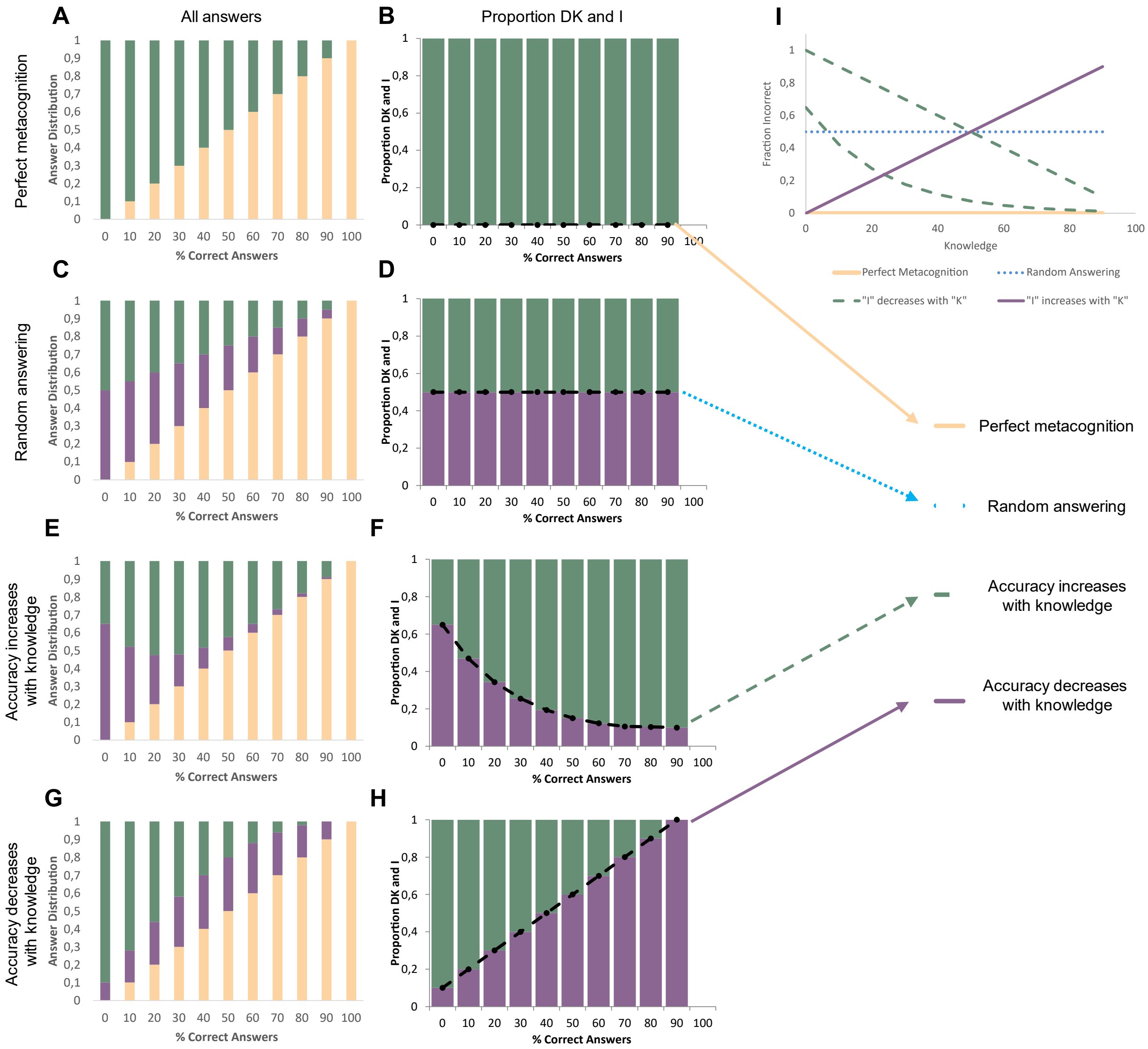

Supplementary Figure 1. Different expectations of the proportions of correct (yellow), incorrect (purple) and “Don’t Know” (green) answers, per knowledge bin **(A,C,E,G)** or proportion of incorrect (purple) and “Don’t Know” (green) within non-correct answers only, per knowledge bin **(B,D,F,H)** depending on different expectations of the relationship between confidence and knowledge **(I)**. Perfect metacognition (A,B, yellow solid line on I) expects all non-correct answers to be of the “Don’t Know” type. Random answering (C,D, dotted blue line on I) expects a constant and even proportion of “Don’t Know” and incorrect answers regardless of knowledge bin. If confidence increases with knowledge (E,F, dashed green lines on I), the proportion of incorrect answers should decrease as knowledge increases. If confidence decreases with knowledge (G,H, solid purple line on I), the proportion of incorrect answers should increase as knowledge increases.

Supplementary Figure 2. EB data

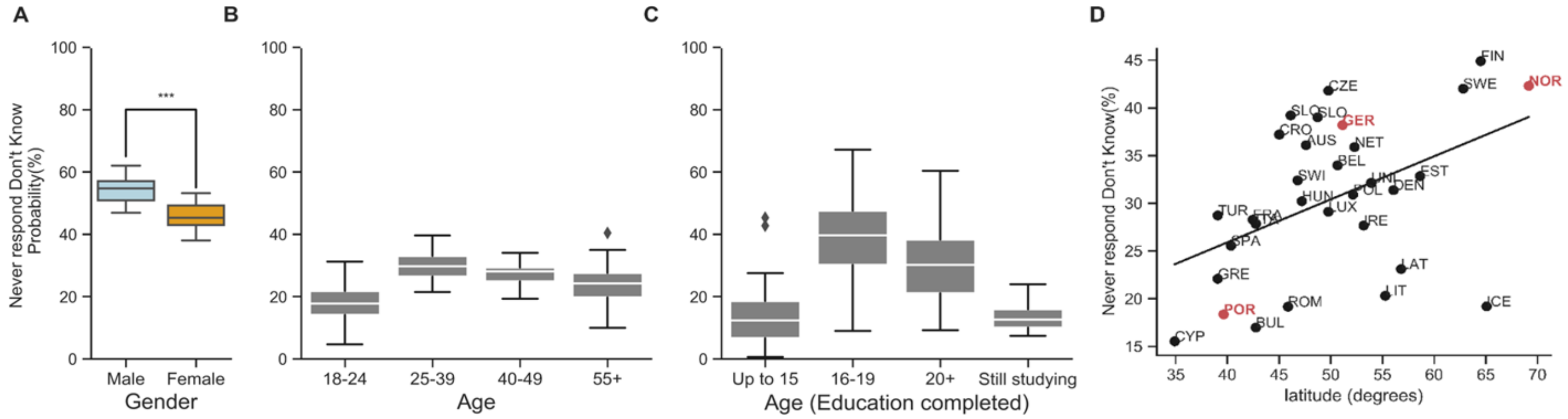

Supplementary Figure 2. EB data. (A) Box plot shows the fraction of female (orange) and male (blue) respondents that never say “Don’t know” across 31 territories. Data was negatively tested for normality using scipy‘s stats module‘s normaltest function ($\alpha$ = 0.001) and for similarity (Mann-Whitney U test). Three black asterisks indicate statistical significance with p-value < 0.001. (B) Box plot shows the fraction of different age group bins that never say “Don’t know” across all 31 territories. Diamonds indicate outliers. A Kruskal-Wallis H-test and all pairwise comparisons were found to be significant with post hoc Tukey’s tests, except 25-39 vs. 40-49 and 40-49 vs. 55+. (C) Box plot shows the fractions of different bins of age at time of completing their education that never say “Don’t know” across all 31 territories. Diamonds mark outliers. A Kruskal-Wallis H-test and all pairwise comparisons were found to be significant with post hoc Tukey’s tests, except Up to 15 vs. Still studying and 16-19 vs. 20+. (D) Scatter plot shows for each territory the fraction of respondents that never say “Don’t know” sorted according to latitude of the territory. Black line shows the linear regression with low correlation represented $R^2$ = 0.21.

Supplementary Figure 3. Lackner survey data

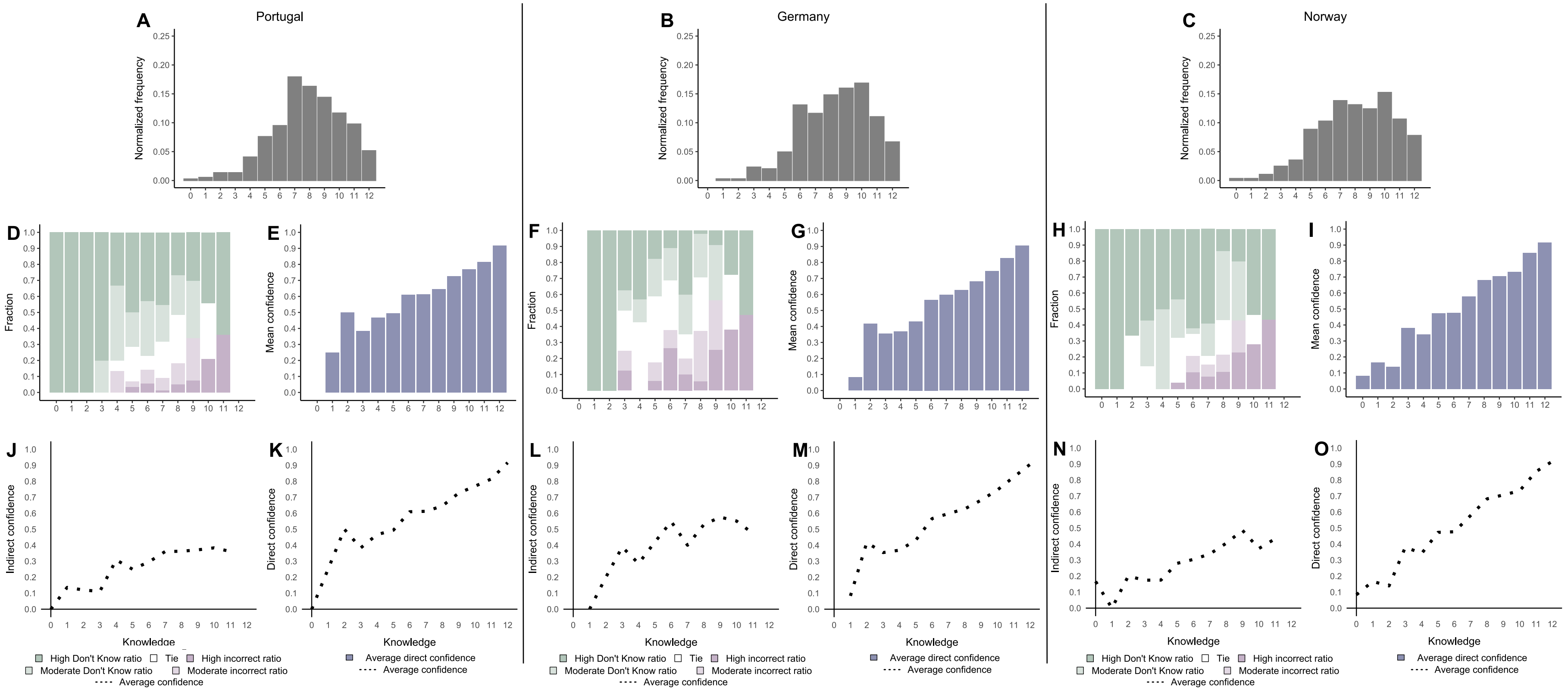

Supplementary Figure 3. Lackner survey data of Portuguese (A,D,E,J,K), German (B,F,G,L,M) and Norwegian (C,H,I,N,O) participants. (A,B,C) knowledge distributions. (D,F,H) stacked bar charts of average fraction of "Don't know" and incorrect answers per knowledge level. (E,G,I) bar chart of average confidence (as measured by percentage of items judged to be correct) per knowledge level. (J,K,L,M,N,O) line chart of average confidence per knowledge level. In (J,L,N), average confidence is calculated as the average proportion of incorrect answers (within all non-correct answers). In (K,M,O), average confidence is calculated as the average number of items judged to be correct.

Supplementary Figure 4. Lackner survey demographics.

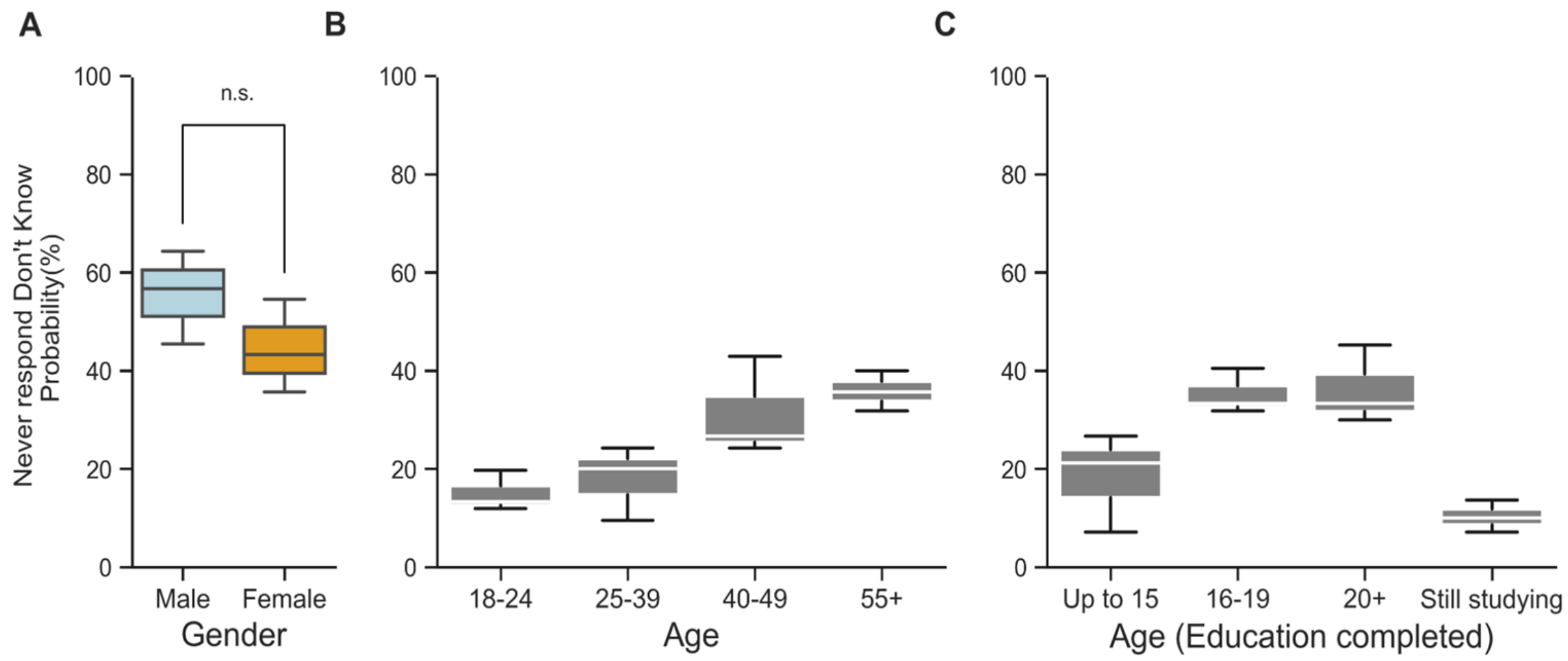

Supplementary Figure 4. Lackner survey demographics. (A) Box plot shows the fraction of female (orange) and male (blue) respondents that never say “Don’t know”. After testing data negatively for normality using scipy‘s stats module‘s normaltest function ($\alpha$ = 0.001) and for similarity using Mann-Whitney U test, no significant difference (n.s.) was found. (B) Box plot shows the fraction of different age group bins that never say “Don’t know”. A Kruskal-Wallis H-test did not provide evidence for difference between means in age groups ($p$ = 0.042). (C) Box plot shows the fractions of different bins of age at time of completing their education that never say “Don’t know”. A Kruskall-Wallis H-test did not find evidence for difference between means in bins of age at time of completing education ($p$ = 0.036).

Supplementary Figure 5. Knowledge distributions for GSS and Pew.

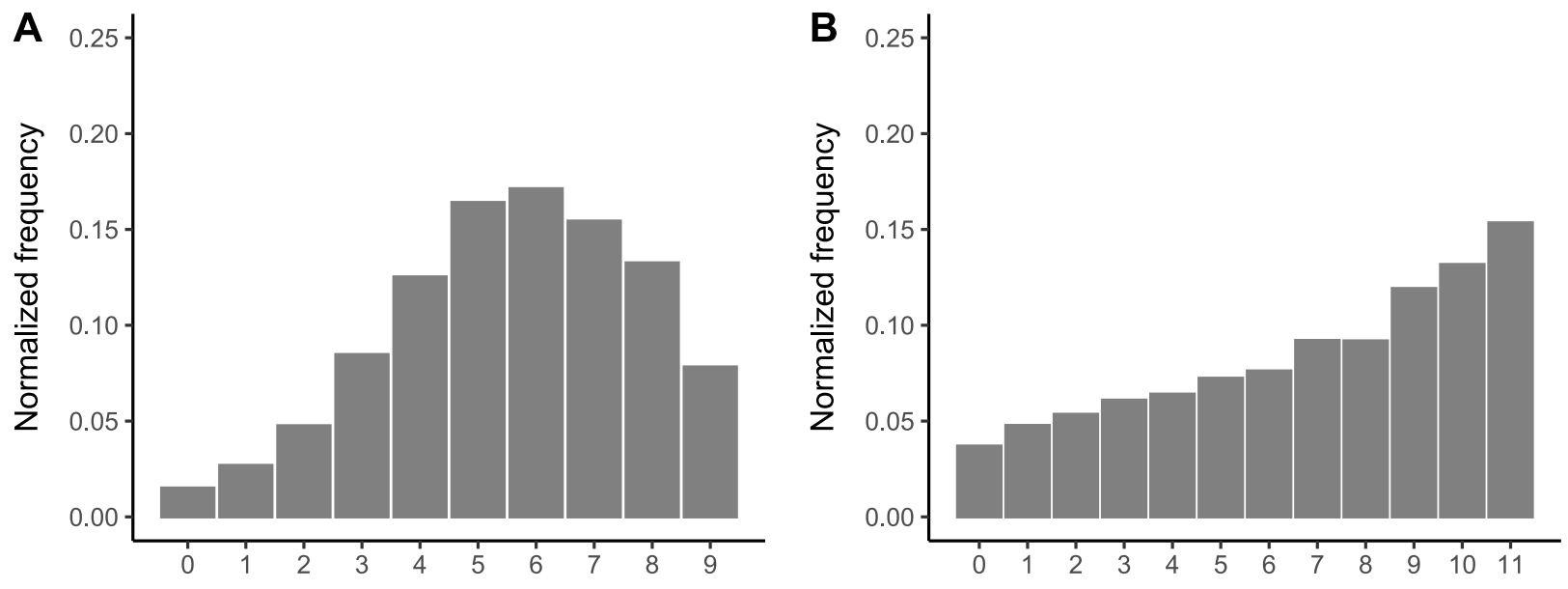

Supplementary Figure 5. Knowledge distributions for GSS (A) and Pew (B). The distributions for the other surveys can be found in the main paper.

Supplementary Figure 6. Alternative calibration models

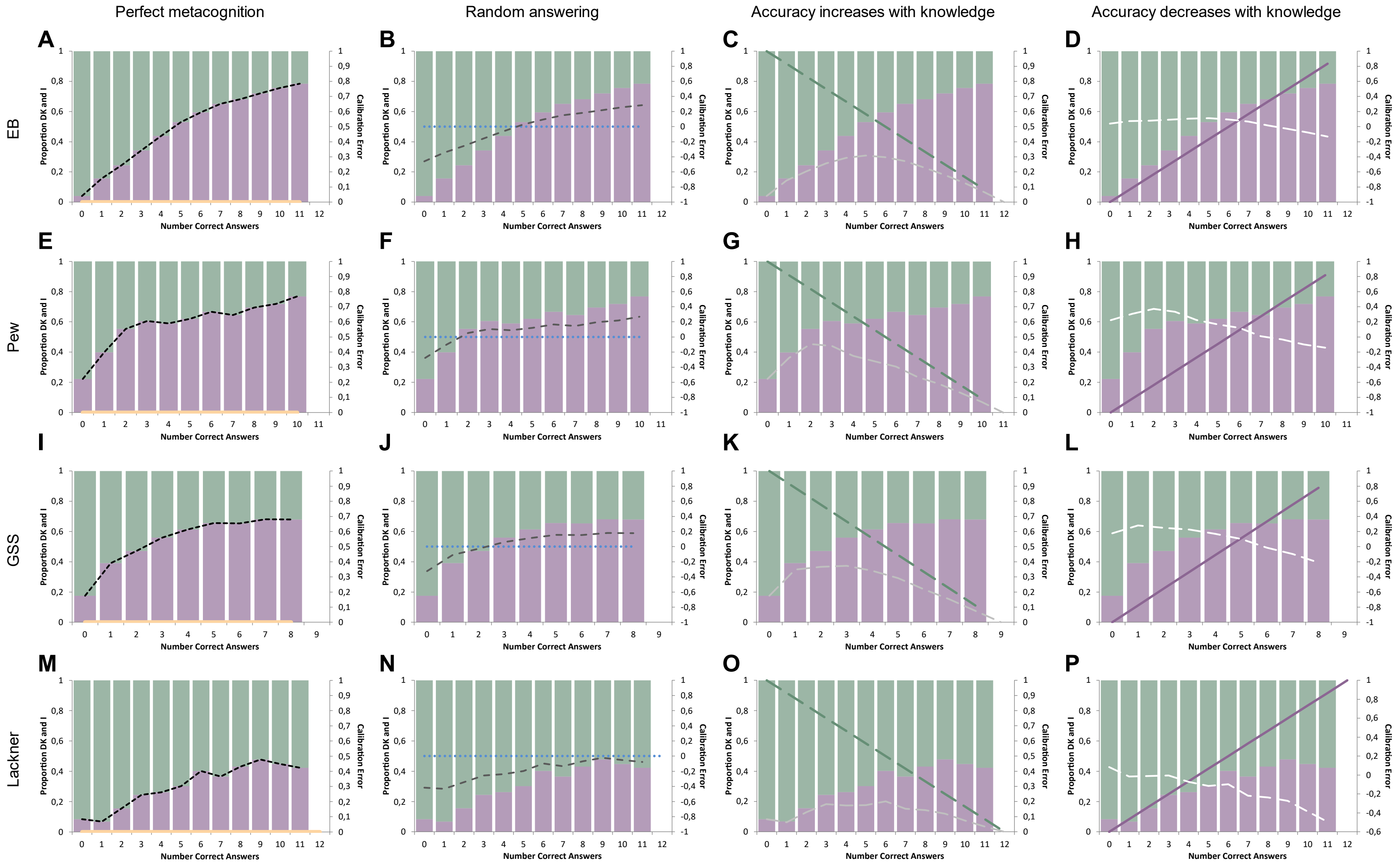

Supplementary Figure 6. Alternative representation of calibration errors, with different null models, with green bars representing actual proportion of “Don’t Know” answers per knowledge bin and purple bars representing actual proportion of incorrect answers per knowledge bin, out of all non-correct answers, for EB (A,B,C,D), Pew (E,F,G,H), GSS (I,J,K,L), and Lackner (M,N,O,P). In (**A,E,I,M**), the null model represents the metacognitive model (yellow solid line), in which any incorrect answer represent a calibration error. The black dashed line represents average calibration error under such representation. In (**B,F,J,N**), the null model represents random guessing (blue dotted line), such that an equal proportion of incorrect and “Don’t Know” answers is expected, regardless of knowledge level. The black dashed line represents average calibration error under such representation. In (**C,G,K,O**), the null model expects confidence to increase in tandem with knowledge (green dashed line). The grey dashed line represents average calibration under such representation. In (**D,H,L,P**), the null model expects confidence to decrease in tandem with knowledge (solid purple line). The white dashed line represents average calibration under such representation.

Supplementary Figure 7. Distribution of the neutral answers to the EB attitudinal questions

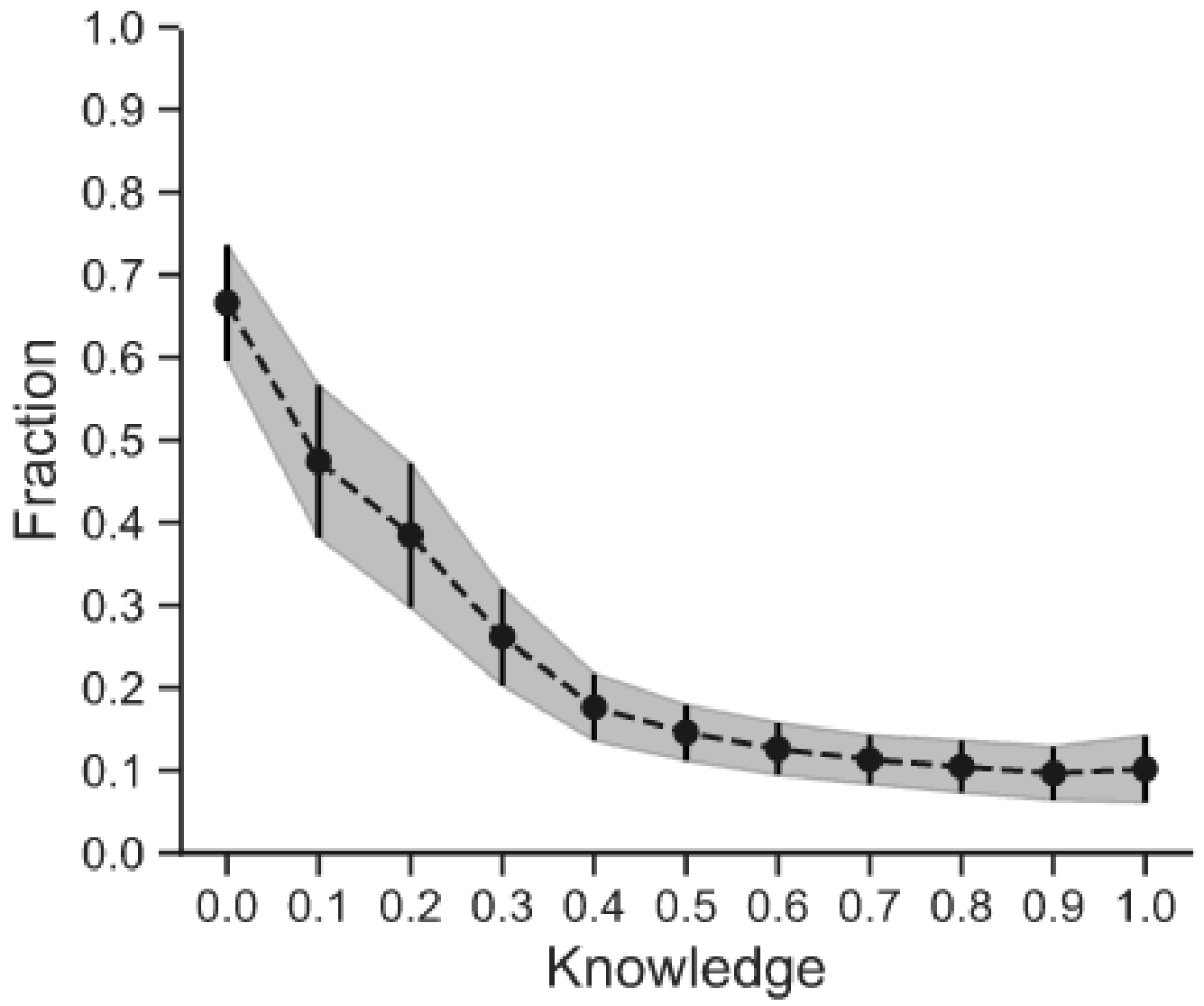

Supplementary Figure 7. EB Neutral answers to the attitudinal questions. Dashed line shows median fraction of neutral responses to all 10 attitude questions combined according to the knowledge-bin (number of questions answered correctly). Black bars and grey edges show standard deviation. The proportion that offers neutral answers (Don’t know or refusing to answer) drops non-linearly as knowledge increases, with individuals in the intermediate knowledge bins offering neutral answers as often as individuals in the higher knowledge bins.

## Supplementary Figure 8. Heatmap of model fits

### 25% guessers

**A**

| M \ SD | 1,21 | 1,71 | 2,21 | 2,71 | 3,21 | 3,71 | 4,21 |
|---|---|---|---|---|---|---|---|
| 7,65 | 3,0E-03 | 1,2E-03 | 6,5E-04 | 6,9E-04 | 1,2E-03 | 1,9E-03 | 2,8E-03 |
| 7,15 | 2,5E-03 | 8,2E-04 | 2,4E-04 | 2,2E-04 | 5,8E-04 | 1,2E-03 | 2,0E-03 |
| 6,65 | 2,4E-03 | 7,3E-04 | 1,2E-04 | **3,8E-05** | 2,9E-04 | 8,1E-04 | 1,5E-03 |
| 6,15 | 2,7E-03 | 9,2E-04 | 2,5E-04 | 8,5E-05 | 2,4E-04 | 6,5E-04 | 1,3E-03 |
| 5,65 | 3,2E-03 | 1,4E-03 | 6,1E-04 | 3,4E-04 | 4,1E-04 | 7,3E-04 | 1,3E-03 |
| 5,15 | 4,0E-03 | 2,0E-03 | 1,1E-03 | 7,6E-04 | 7,4E-04 | 9,7E-04 | 1,4E-03 |
| 4,65 | 5,0E-03 | 2,8E-03 | 1,8E-03 | 1,3E-03 | 1,2E-03 | 1,4E-03 | 1,9E-03 |
| 4,15 | 6,1E-03 | 3,7E-03 | 2,6E-03 | 2,0E-03 | 1,9E-03 | 2,1E-03 | 2,4E-03 |
| 3,65 | 7,3E-03 | 4,7E-03 | 3,4E-03 | 2,9E-03 | 2,8E-03 | 2,9E-03 | 3,2E-03 |
| 3,15 | 8,4E-03 | 5,7E-03 | 4,4E-03 | 3,9E-03 | 3,8E-03 | 4,0E-03 | 4,3E-03 |
| 2,65 | 9,5E-03 | 6,7E-03 | 5,6E-03 | 5,1E-03 | 5,0E-03 | 5,2E-03 | 5,4E-03 |

### 50% guessers

**B**

| M \ SD | 1,21 | 1,71 | 2,21 | 2,71 | 3,21 | 3,71 | 4,21 |
|---|---|---|---|---|---|---|---|
| 7,65 | 2,8E-03 | 2,1E-03 | 1,8E-03 | 1,9E-03 | 2,3E-03 | 2,9E-03 | 3,6E-03 |
| 7,15 | 1,8E-03 | 1,2E-03 | 1,0E-03 | 1,1E-03 | 1,4E-03 | 1,9E-03 | 2,6E-03 |
| 6,65 | 1,1E-03 | 6,6E-04 | 5,0E-04 | 5,4E-04 | 7,9E-04 | 1,2E-03 | 1,8E-03 |
| 6,15 | 7,5E-04 | 3,1E-04 | 1,8E-04 | 2,1E-04 | 3,8E-04 | 7,5E-04 | 1,3E-03 |
| 5,65 | 6,7E-04 | 2,0E-04 | 5,5E-05 | 5,5E-05 | 1,9E-04 | 4,7E-04 | 9,1E-04 |
| 5,15 | 8,3E-04 | 2,8E-04 | 9,9E-05 | 5,8E-05 | 1,5E-04 | 3,7E-04 | 7,4E-04 |
| 4,65 | 1,2E-03 | 5,2E-04 | 2,8E-04 | 1,9E-04 | 2,4E-04 | 4,2E-04 | 7,5E-04 |
| 4,15 | 1,7E-03 | 9,1E-04 | 5,7E-04 | 4,5E-04 | 4,6E-04 | 6,3E-04 | 8,9E-04 |
| 3,65 | 2,3E-03 | 1,4E-03 | 9,5E-04 | 8,1E-04 | 8,3E-04 | 9,6E-04 | 1,2E-03 |
| 3,15 | 2,9E-03 | 1,9E-03 | 1,4E-03 | 1,3E-03 | 1,3E-03 | 1,4E-03 | 1,6E-03 |
| 2,65 | 3,5E-03 | 2,5E-03 | 2,0E-03 | 1,9E-03 | 1,9E-03 | 2,0E-03 | 2,1E-03 |

### 75% guessers

**C**

| M \ SD | 1,21 | 1,71 | 2,21 | 2,71 | 3,21 | 3,71 | 4,21 |
|---|---|---|---|---|---|---|---|
| 7,65 | 5,1E-03 | 4,5E-03 | 4,0E-03 | 3,8E-03 | 4,0E-03 | 4,3E-03 | 4,8E-03 |
| 7,15 | 3,7E-03 | 3,3E-03 | 3,0E-03 | 2,7E-03 | 2,8E-03 | 3,1E-03 | 3,5E-03 |
| 6,65 | 2,6E-03 | 2,4E-03 | 2,1E-03 | 1,9E-03 | 2,0E-03 | 2,1E-03 | 2,5E-03 |
| 6,15 | 1,8E-03 | 1,6E-03 | 1,4E-03 | 1,3E-03 | 1,3E-03 | 1,4E-03 | 1,8E-03 |
| 5,65 | 1,2E-03 | 1,1E-03 | 9,4E-04 | 8,2E-04 | 7,8E-04 | 8,9E-04 | 1,2E-03 |
| 5,15 | 8,7E-04 | 7,5E-04 | 6,1E-04 | 4,9E-04 | 4,6E-04 | 5,4E-04 | 7,5E-04 |
| 4,65 | 6,9E-04 | 5,1E-04 | 4,0E-04 | 3,0E-04 | 2,5E-04 | 3,0E-04 | 4,5E-04 |
| 4,15 | 6,4E-04 | 4,5E-04 | 3,2E-04 | 2,1E-04 | 1,5E-04 | 1,8E-04 | 3,1E-04 |
| 3,65 | 7,3E-04 | 4,8E-04 | 3,3E-04 | 2,3E-04 | 1,7E-04 | 1,7E-04 | 2,4E-04 |
| 3,15 | 9,1E-04 | 6,1E-04 | 4,4E-04 | 3,1E-04 | 2,6E-04 | 2,4E-04 | 2,8E-04 |
| 2,65 | 1,1E-03 | 8,2E-04 | 6,2E-04 | 5,1E-04 | 4,3E-04 | 3,9E-04 | 4,0E-04 |

**D**

| M \ SD | 0,64 | 1,14 | 1,64 | 2,14 | 2,64 | 3,14 | 3,64 |
|---|---|---|---|---|---|---|---|
| 6,07 | 1,2E-02 | 3,9E-03 | 1,5E-03 | 1,3E-03 | 2,2E-03 | 3,6E-03 | 5,2E-03 |
| 5,57 | 9,7E-03 | 2,9E-03 | 6,5E-04 | 2,7E-04 | 9,1E-04 | 2,1E-03 | 3,6E-03 |
| 5,07 | 1,0E-02 | 2,8E-03 | 5,8E-04 | **3,6E-05** | 4,0E-04 | 1,4E-03 | 2,7E-03 |
| 4,57 | 1,0E-02 | 3,5E-03 | 1,2E-03 | 3,9E-04 | 5,2E-04 | 1,3E-03 | 2,4E-03 |
| 4,07 | 1,3E-02 | 4,9E-03 | 2,3E-03 | 1,2E-03 | 1,2E-03 | 1,7E-03 | 2,7E-03 |
| 3,57 | 1,5E-02 | 6,7E-03 | 3,7E-03 | 2,5E-03 | 2,2E-03 | 2,5E-03 | 3,4E-03 |
| 3,07 | 1,8E-02 | 8,8E-03 | 5,5E-03 | 4,0E-03 | 3,6E-03 | 3,9E-03 | 4,6E-03 |
| 2,57 | 1,9E-02 | 1,1E-02 | 7,4E-03 | 5,9E-03 | 5,5E-03 | 5,7E-03 | 6,2E-03 |
| 2,07 | 2,2E-02 | 1,3E-02 | 9,6E-03 | 8,4E-03 | 8,0E-03 | 8,1E-03 | 8,3E-03 |
| 1,57 | 2,4E-02 | 1,5E-02 | 1,2E-02 | 1,1E-02 | 1,1E-02 | 1,1E-02 | 1,1E-02 |
| 1,07 | 2,6E-02 | 1,8E-02 | 1,6E-02 | 1,5E-02 | 1,4E-02 | 1,4E-02 | 1,3E-02 |

**E**

| M \ SD | 0,64 | 1,14 | 1,64 | 2,14 | 2,64 | 3,14 | 3,64 |
|---|---|---|---|---|---|---|---|
| 6,07 | 8,1E-03 | 4,5E-03 | 3,2E-03 | 3,2E-03 | 4,0E-03 | 5,0E-03 | 6,4E-03 |
| 5,57 | 5,2E-03 | 2,6E-03 | 1,6E-03 | 1,4E-03 | 2,0E-03 | 3,0E-03 | 4,2E-03 |
| 5,07 | 4,0E-03 | 1,5E-03 | 6,7E-04 | 5,0E-04 | 9,1E-04 | 1,7E-03 | 2,8E-03 |
| 4,57 | 3,4E-03 | 1,0E-03 | 3,0E-04 | 1,1E-04 | 3,3E-04 | 9,7E-04 | 1,9E-03 |
| 4,07 | 4,0E-03 | 1,2E-03 | 4,4E-04 | 1,4E-04 | 2,4E-04 | 6,9E-04 | 1,5E-03 |
| 3,57 | 4,7E-03 | 1,9E-03 | 9,2E-04 | 5,3E-04 | 5,1E-04 | 8,1E-04 | 1,4E-03 |
| 3,07 | 6,2E-03 | 2,8E-03 | 1,7E-03 | 1,2E-03 | 1,1E-03 | 1,3E-03 | 1,8E-03 |
| 2,57 | 7,3E-03 | 4,0E-03 | 2,7E-03 | 2,1E-03 | 2,0E-03 | 2,1E-03 | 2,4E-03 |
| 2,07 | 9,0E-03 | 5,2E-03 | 3,8E-03 | 3,4E-03 | 3,2E-03 | 3,2E-03 | 3,3E-03 |
| 1,57 | 1,0E-02 | 6,5E-03 | 5,3E-03 | 4,8E-03 | 4,6E-03 | 4,5E-03 | 4,5E-03 |
| 1,07 | 1,1E-02 | 8,1E-03 | 7,2E-03 | 6,7E-03 | 6,3E-03 | 6,0E-03 | 5,7E-03 |

**F**

| M \ SD | 0,64 | 1,14 | 1,64 | 2,14 | 2,64 | 3,14 | 3,64 |
|---|---|---|---|---|---|---|---|
| 6,07 | 9,3E-03 | 7,6E-03 | 6,4E-03 | 6,1E-03 | 6,5E-03 | 7,0E-03 | 8,1E-03 |
| 5,57 | 6,1E-03 | 5,1E-03 | 4,2E-03 | 3,7E-03 | 4,0E-03 | 4,6E-03 | 5,4E-03 |
| 5,07 | 3,9E-03 | 3,3E-03 | 2,7E-03 | 2,2E-03 | 2,3E-03 | 2,8E-03 | 3,6E-03 |
| 4,57 | 2,5E-03 | 2,0E-03 | 1,6E-03 | 1,2E-03 | 1,2E-03 | 1,6E-03 | 2,2E-03 |
| 4,07 | 1,8E-03 | 1,3E-03 | 9,7E-04 | 6,2E-04 | 5,2E-04 | 7,5E-04 | 1,3E-03 |
| 3,57 | 1,5E-03 | 9,6E-04 | 6,4E-04 | 3,8E-04 | 2,3E-04 | 3,4E-04 | 7,0E-04 |
| 3,07 | 1,7E-03 | 9,9E-04 | 6,8E-04 | 3,6E-04 | 2,0E-04 | 2,2E-04 | 4,3E-04 |
| 2,57 | 2,0E-03 | 1,3E-03 | 8,9E-04 | 6,1E-04 | 4,0E-04 | 3,3E-04 | 4,3E-04 |
| 2,07 | 2,6E-03 | 1,7E-03 | 1,3E-03 | 1,1E-03 | 8,1E-04 | 6,7E-04 | 6,4E-04 |
| 1,57 | 3,2E-03 | 2,3E-03 | 1,9E-03 | 1,6E-03 | 1,4E-03 | 1,1E-03 | 1,0E-03 |
| 1,07 | 4,0E-03 | 3,1E-03 | 2,7E-03 | 2,4E-03 | 2,0E-03 | 1,7E-03 | 1,5E-03 |

**G**

| $b_0$ \ $b_1$ | 0,000 | 0,005 | 0,010 | 0,015 | 0,020 | 0,025 | 0,030 |
|---|---|---|---|---|---|---|---|
| 0,006 | 7,8E-04 | 2,9E-04 | 4,4E-04 | 4,9E-04 | 5,3E-04 | 5,5E-04 | 5,6E-04 |
| 0,011 | 7,7E-04 | 1,6E-04 | 3,1E-04 | 4,0E-04 | 4,5E-04 | 4,8E-04 | 5,0E-04 |
| 0,016 | 7,8E-04 | 1,1E-04 | 2,3E-04 | 3,2E-04 | 3,8E-04 | 4,2E-04 | 4,5E-04 |
| 0,021 | 7,8E-04 | 8,9E-05 | 1,8E-04 | 2,6E-04 | 3,3E-04 | 3,7E-04 | 4,1E-04 |
| 0,026 | 7,8E-04 | 8,9E-05 | 1,4E-04 | 2,2E-04 | 2,8E-04 | 3,3E-04 | 3,6E-04 |
| 0,031 | 7,8E-04 | 1,0E-04 | 1,1E-04 | 1,8E-04 | 2,4E-04 | 2,9E-04 | 3,3E-04 |
| 0,036 | 7,8E-04 | 1,2E-04 | 9,9E-05 | 1,5E-04 | 2,1E-04 | 2,6E-04 | 3,0E-04 |
| 0,041 | 7,8E-04 | 1,4E-04 | 9,0E-05 | 1,3E-04 | 1,8E-04 | 2,3E-04 | 2,7E-04 |
| 0,046 | 7,8E-04 | 1,6E-04 | **8,7E-05** | 1,2E-04 | 1,6E-04 | 2,1E-04 | 2,5E-04 |
| 0,051 | 7,8E-04 | 1,8E-04 | 8,8E-05 | 1,0E-04 | 1,4E-04 | 1,9E-04 | 2,2E-04 |
| 0,056 | 7,8E-04 | 2,0E-04 | 9,0E-05 | 9,6E-05 | 1,3E-04 | 1,7E-04 | 2,0E-04 |

**H**

| $b_0$ \ $b_1$ | 0,000 | 0,005 | 0,010 | 0,015 | 0,020 | 0,025 | 0,030 |
|---|---|---|---|---|---|---|---|
| 0,006 | 5,6E-04 | 4,9E-04 | 6,7E-04 | 7,4E-04 | 7,8E-04 | 8,0E-04 | 8,2E-04 |
| 0,011 | 5,6E-04 | 3,1E-04 | 5,2E-04 | 6,2E-04 | 6,8E-04 | 7,2E-04 | 7,5E-04 |
| 0,016 | 5,7E-04 | 2,1E-04 | 4,1E-04 | 5,3E-04 | 6,0E-04 | 6,5E-04 | 6,9E-04 |
| 0,021 | 5,7E-04 | 1,7E-04 | 3,3E-04 | 4,5E-04 | 5,3E-04 | 5,9E-04 | 6,3E-04 |
| 0,026 | 5,7E-04 | 1,4E-04 | 2,7E-04 | 3,9E-04 | 4,8E-04 | 5,3E-04 | 5,8E-04 |
| 0,031 | 5,7E-04 | 1,3E-04 | 2,3E-04 | 3,4E-04 | 4,2E-04 | 4,9E-04 | 5,4E-04 |
| 0,036 | 5,7E-04 | 1,3E-04 | 2,0E-04 | 3,0E-04 | 3,8E-04 | 4,5E-04 | 5,0E-04 |
| 0,041 | 5,7E-04 | 1,4E-04 | 1,7E-04 | 2,6E-04 | 3,4E-04 | 4,1E-04 | 4,6E-04 |
| 0,046 | 5,7E-04 | 1,5E-04 | 1,6E-04 | 2,3E-04 | 3,1E-04 | 3,7E-04 | 4,3E-04 |
| 0,051 | 5,7E-04 | 1,5E-04 | 1,5E-04 | 2,1E-04 | 2,8E-04 | 3,4E-04 | 3,9E-04 |
| 0,056 | 5,7E-04 | 1,6E-04 | 1,4E-04 | 1,9E-04 | 2,6E-04 | 3,2E-04 | 3,7E-04 |

**I**

| $b_0$ \ $b_1$ | 0,000 | 0,005 | 0,010 | 0,015 | 0,020 | 0,025 | 0,030 |
|---|---|---|---|---|---|---|---|
| 0,006 | 5,4E-04 | 7,6E-04 | 9,8E-04 | 1,1E-03 | 1,1E-03 | 1,1E-03 | 1,2E-03 |
| 0,011 | 5,3E-04 | 5,3E-04 | 7,9E-04 | 9,2E-04 | 9,9E-04 | 1,0E-03 | 1,1E-03 |
| 0,016 | 5,4E-04 | 4,0E-04 | 6,6E-04 | 8,1E-04 | 9,0E-04 | 9,5E-04 | 9,9E-04 |
| 0,021 | 5,4E-04 | 3,3E-04 | 5,6E-04 | 7,2E-04 | 8,1E-04 | 8,8E-04 | 9,4E-04 |
| 0,026 | 5,4E-04 | 2,8E-04 | 4,9E-04 | 6,4E-04 | 7,4E-04 | 8,1E-04 | 8,6E-04 |
| 0,031 | 5,4E-04 | 2,6E-04 | 4,2E-04 | 5,7E-04 | 6,7E-04 | 7,6E-04 | 8,2E-04 |
| 0,036 | 5,4E-04 | 2,4E-04 | 3,8E-04 | 5,2E-04 | 6,2E-04 | 7,1E-04 | 7,7E-04 |
| 0,041 | 5,4E-04 | 2,4E-04 | 3,5E-04 | 4,7E-04 | 5,7E-04 | 6,6E-04 | 7,2E-04 |
| 0,046 | 5,5E-04 | 2,4E-04 | 3,2E-04 | 4,3E-04 | 5,3E-04 | 6,1E-04 | 6,8E-04 |
| 0,051 | 5,4E-04 | 2,3E-04 | 2,9E-04 | 4,0E-04 | 5,0E-04 | 5,8E-04 | 6,4E-04 |
| 0,056 | 5,4E-04 | 2,4E-04 | 2,8E-04 | 3,7E-04 | 4,6E-04 | 5,4E-04 | 6,1E-04 |

**J**

| M \ SD | 0,86 | 1,36 | 1,86 | 2,36 | 2,86 | 3,36 | 3,86 |
|---|---|---|---|---|---|---|---|
| 8,6 | 6,2E-03 | 2,3E-03 | 1,0E-03 | 9,1E-04 | 1,5E-03 | 2,4E-03 | 3,5E-03 |
| 8,1 | 5,3E-03 | 1,7E-03 | 4,2E-04 | 2,4E-04 | 6,7E-04 | 1,4E-03 | 2,4E-03 |
| 7,6 | 4,8E-03 | 1,4E-03 | 2,9E-04 | **3,1E-05** | 3,0E-04 | 9,2E-04 | 1,7E-03 |
| 7,1 | 4,8E-03 | 1,6E-03 | 5,1E-04 | 1,7E-04 | 2,9E-04 | 7,3E-04 | 1,4E-03 |
| 6,6 | 5,5E-03 | 2,3E-03 | 1,1E-03 | 6,1E-04 | 5,7E-04 | 8,2E-04 | 1,3E-03 |
| 6,1 | 6,8E-03 | 3,3E-03 | 1,9E-03 | 1,3E-03 | 1,1E-03 | 1,1E-03 | 1,5E-03 |
| 5,6 | 8,5E-03 | 4,6E-03 | 3,0E-03 | 2,1E-03 | 1,7E-03 | 1,6E-03 | 1,9E-03 |
| 5,1 | 1,0E-02 | 6,0E-03 | 4,1E-03 | 3,0E-03 | 2,5E-03 | 2,3E-03 | 2,4E-03 |
| 4,6 | 1,2E-02 | 7,5E-03 | 5,3E-03 | 4,0E-03 | 3,4E-03 | 3,1E-03 | 3,2E-03 |
| 4,1 | 1,4E-02 | 8,9E-03 | 6,5E-03 | 5,1E-03 | 4,4E-03 | 4,1E-03 | 4,1E-03 |
| 3,6 | 1,5E-02 | 1,0E-02 | 7,6E-03 | 6,2E-03 | 5,5E-03 | 5,2E-03 | 5,1E-03 |

**K**

| M \ SD | 0,86 | 1,36 | 1,86 | 2,36 | 2,86 | 3,36 | 3,86 |
|---|---|---|---|---|---|---|---|
| 8,6 | 4,8E-03 | 3,0E-03 | 2,3E-03 | 2,3E-03 | 2,8E-03 | 3,5E-03 | 4,4E-03 |
| 8,1 | 3,2E-03 | 1,8E-03 | 1,2E-03 | 1,2E-03 | 1,5E-03 | 2,2E-03 | 2,9E-03 |
| 7,6 | 2,1E-03 | 9,7E-04 | 5,6E-04 | 5,0E-04 | 7,7E-04 | 1,3E-03 | 2,0E-03 |
| 7,1 | 1,5E-03 | 5,3E-04 | 2,3E-04 | 1,8E-04 | 3,4E-04 | 7,5E-04 | 1,3E-03 |
| 6,6 | 1,4E-03 | 4,6E-04 | 2,0E-04 | 9,8E-05 | 1,9E-04 | 4,7E-04 | 9,3E-04 |
| 6,1 | 1,8E-03 | 7,3E-04 | 3,9E-04 | 2,5E-04 | 2,5E-04 | 4,1E-04 | 7,4E-04 |
| 5,6 | 2,6E-03 | 1,3E-03 | 8,1E-04 | 5,5E-04 | 4,6E-04 | 5,3E-04 | 7,8E-04 |
| 5,1 | 3,6E-03 | 2,0E-03 | 1,3E-03 | 1,0E-03 | 8,2E-04 | 8,2E-04 | 9,5E-04 |
| 4,6 | 4,6E-03 | 2,8E-03 | 2,0E-03 | 1,5E-03 | 1,3E-03 | 1,2E-03 | 1,3E-03 |
| 4,1 | 5,6E-03 | 3,6E-03 | 2,7E-03 | 2,1E-03 | 1,8E-03 | 1,7E-03 | 1,7E-03 |
| 3,6 | 6,4E-03 | 4,3E-03 | 3,3E-03 | 2,8E-03 | 2,4E-03 | 2,3E-03 | 2,3E-03 |

**L**

| M \ SD | 0,86 | 1,36 | 1,86 | 2,36 | 2,86 | 3,36 | 3,86 |
|---|---|---|---|---|---|---|---|
| 8,6 | 6,4E-03 | 5,4E-03 | 4,6E-03 | 4,4E-03 | 4,6E-03 | 5,0E-03 | 5,7E-03 |
| 8,1 | 4,6E-03 | 3,8E-03 | 3,2E-03 | 2,9E-03 | 3,0E-03 | 3,4E-03 | 3,9E-03 |
| 7,6 | 3,0E-03 | 2,6E-03 | 2,2E-03 | 1,9E-03 | 1,9E-03 | 2,2E-03 | 2,7E-03 |
| 7,1 | 2,0E-03 | 1,7E-03 | 1,4E-03 | 1,2E-03 | 1,1E-03 | 1,4E-03 | 1,7E-03 |
| 6,6 | 1,3E-03 | 1,1E-03 | 9,3E-04 | 6,9E-04 | 6,4E-04 | 7,5E-04 | 1,1E-03 |
| 6,1 | 1,1E-03 | 8,0E-04 | 6,4E-04 | 4,4E-04 | 3,3E-04 | 3,9E-04 | 6,0E-04 |
| 5,6 | 9,9E-04 | 7,4E-04 | 5,1E-04 | 3,3E-04 | 2,0E-04 | 2,0E-04 | 3,4E-04 |
| 5,1 | 1,2E-03 | 8,1E-04 | 5,8E-04 | 3,8E-04 | 2,1E-04 | 1,7E-04 | 2,5E-04 |
| 4,6 | 1,5E-03 | 1,0E-03 | 7,2E-04 | 5,2E-04 | 3,5E-04 | 2,4E-04 | 2,7E-04 |
| 4,1 | 1,9E-03 | 1,3E-03 | 1,0E-03 | 7,5E-04 | 5,6E-04 | 4,4E-04 | 4,2E-04 |
| 3,6 | 2,3E-03 | 1,7E-03 | 1,3E-03 | 1,1E-03 | 8,4E-04 | 7,2E-04 | 6,3E-04 |

**M**

| M \ SD | 0,9 | 1,4 | 1,9 | 2,4 | 2,9 | 3,4 | 3,9 |
|---|---|---|---|---|---|---|---|
| 5,91 | 6,0E-03 | 2,4E-03 | 1,3E-03 | 1,2E-03 | 1,7E-03 | 2,6E-03 | 3,9E-03 |
| 5,41 | 4,5E-03 | 1,4E-03 | 5,1E-04 | 4,3E-04 | 8,8E-04 | 1,7E-03 | 2,8E-03 |
| 4,91 | 4,0E-03 | 1,1E-03 | 2,2E-04 | **1,4E-04** | 5,1E-04 | 1,3E-03 | 2,3E-03 |
| 4,41 | 4,5E-03 | 1,4E-03 | 3,9E-04 | 2,1E-04 | 5,1E-04 | 1,2E-03 | 2,2E-03 |
| 3,91 | 6,0E-03 | 2,2E-03 | 9,6E-04 | 6,3E-04 | 8,9E-04 | 1,5E-03 | 2,4E-03 |
| 3,41 | 7,8E-03 | 3,4E-03 | 1,8E-03 | 1,4E-03 | 1,6E-03 | 2,1E-03 | 3,0E-03 |
| 2,91 | 9,8E-03 | 4,9E-03 | 3,0E-03 | 2,5E-03 | 2,7E-03 | 3,2E-03 | 4,0E-03 |
| 2,41 | 1,1E-02 | 6,4E-03 | 4,5E-03 | 4,0E-03 | 4,2E-03 | 4,7E-03 | 5,3E-03 |
| 1,91 | 1,3E-02 | 8,2E-03 | 6,5E-03 | 6,1E-03 | 6,2E-03 | 6,6E-03 | 7,0E-03 |
| 1,41 | 1,4E-02 | 1,0E-02 | 9,1E-03 | 8,7E-03 | 8,7E-03 | 8,8E-03 | 9,0E-03 |
| 0,91 | 1,7E-02 | 1,4E-02 | 1,3E-02 | 1,2E-02 | 1,2E-02 | 1,1E-02 | 1,1E-02 |

**N**

| M \ SD | 0,9 | 1,4 | 1,9 | 2,4 | 2,9 | 3,4 | 3,9 |
|---|---|---|---|---|---|---|---|
| 5,91 | 4,9E-03 | 3,6E-03 | 3,1E-03 | 3,0E-03 | 3,4E-03 | 4,1E-03 | 5,1E-03 |
| 5,41 | 2,9E-03 | 2,1E-03 | 1,8E-03 | 1,8E-03 | 2,2E-03 | 2,8E-03 | 3,6E-03 |
| 4,91 | 1,7E-03 | 1,0E-03 | 9,4E-04 | 1,0E-03 | 1,3E-03 | 1,8E-03 | 2,6E-03 |
| 4,41 | 1,2E-03 | 4,9E-04 | 4,0E-04 | 5,2E-04 | 7,7E-04 | 1,3E-03 | 2,0E-03 |
| 3,91 | 1,5E-03 | 4,2E-04 | 2,3E-04 | 2,9E-04 | 5,4E-04 | 9,8E-04 | 1,6E-03 |
| 3,41 | 2,2E-03 | 6,9E-04 | 3,2E-04 | 3,3E-04 | 5,6E-04 | 9,5E-04 | 1,5E-03 |
| 2,91 | 3,0E-03 | 1,2E-03 | 6,7E-04 | 6,2E-04 | 8,3E-04 | 1,2E-03 | 1,7E-03 |
| 2,41 | 3,8E-03 | 1,8E-03 | 1,2E-03 | 1,2E-03 | 1,3E-03 | 1,7E-03 | 2,1E-03 |
| 1,91 | 4,6E-03 | 2,6E-03 | 2,0E-03 | 2,0E-03 | 2,1E-03 | 2,4E-03 | 2,7E-03 |
| 1,41 | 5,2E-03 | 3,6E-03 | 3,2E-03 | 3,1E-03 | 3,2E-03 | 3,3E-03 | 3,4E-03 |
| 0,91 | 6,4E-03 | 5,2E-03 | 4,8E-03 | 4,5E-03 | 4,4E-03 | 4,4E-03 | 4,4E-03 |

**O**

| M \ SD | 0,9 | 1,4 | 1,9 | 2,4 | 2,9 | 3,4 | 3,9 |
|---|---|---|---|---|---|---|---|
| 5,91 | 7,7E-03 | 7,0E-03 | 6,3E-03 | 5,8E-03 | 5,8E-03 | 6,2E-03 | 6,8E-03 |
| 5,41 | 5,5E-03 | 5,1E-03 | 4,7E-03 | 4,3E-03 | 4,3E-03 | 4,5E-03 | 5,0E-03 |
| 4,91 | 3,7E-03 | 3,7E-03 | 3,4E-03 | 3,1E-03 | 3,0E-03 | 3,2E-03 | 3,7E-03 |
| 4,41 | 2,6E-03 | 2,5E-03 | 2,3E-03 | 2,2E-03 | 2,1E-03 | 2,3E-03 | 2,6E-03 |
| 3,91 | 1,8E-03 | 1,7E-03 | 1,6E-03 | 1,5E-03 | 1,4E-03 | 1,5E-03 | 1,9E-03 |
| 3,41 | 1,5E-03 | 1,2E-03 | 1,0E-03 | 9,6E-04 | 9,2E-04 | 1,1E-03 | 1,3E-03 |
| 2,91 | 1,2E-03 | 8,2E-04 | 6,9E-04 | 6,4E-04 | 6,6E-04 | 7,4E-04 | 9,3E-04 |
| 2,41 | 1,1E-03 | 6,8E-04 | 5,2E-04 | 5,0E-04 | 5,2E-04 | 5,8E-04 | 7,6E-04 |
| 1,91 | 1,2E-03 | 6,7E-04 | 5,6E-04 | 5,1E-04 | 5,3E-04 | 5,9E-04 | 6,8E-04 |
| 1,41 | 1,2E-03 | 8,4E-04 | 7,3E-04 | 7,0E-04 | 6,8E-04 | 6,9E-04 | 7,4E-04 |
| 0,91 | 1,5E-03 | 1,2E-03 | 1,1E-03 | 1,0E-03 | 9,6E-04 | 9,3E-04 | 9,0E-04 |

Supplementary Figure 8. Heatmap of model fits (MSE in scientific notation) for the parameter space for EB (A,B,C), GSS (D,E,F), Pew (G,H,I), Lackner (J,K,L) and Fernbach (M,N,O). EB, GSS, Lackner and Fernback were fit using a Normal distribution (with mean M and standard deviation SD) and Pew was fit using a Normal and a linear distribution (with intercept $b_0$ intercept and slope $b_1$). In the case of Pew only linear is shown as the fit for the normal distribution was very poor. The percentage of estimated guessers is 25% in (A,D,G,J,M), 50% in (B,E,H,J,N) , and 75% in (C,F,I,L,O). Bold and black sqares marks the best fitting parameters, which always occurred when the percentage of guessers was 25%, and dark grey squares mark the best fit for the 75% guessers (see Methods and Supplementary Figure 9).

Supplementary Figure 9. Simulation best fits

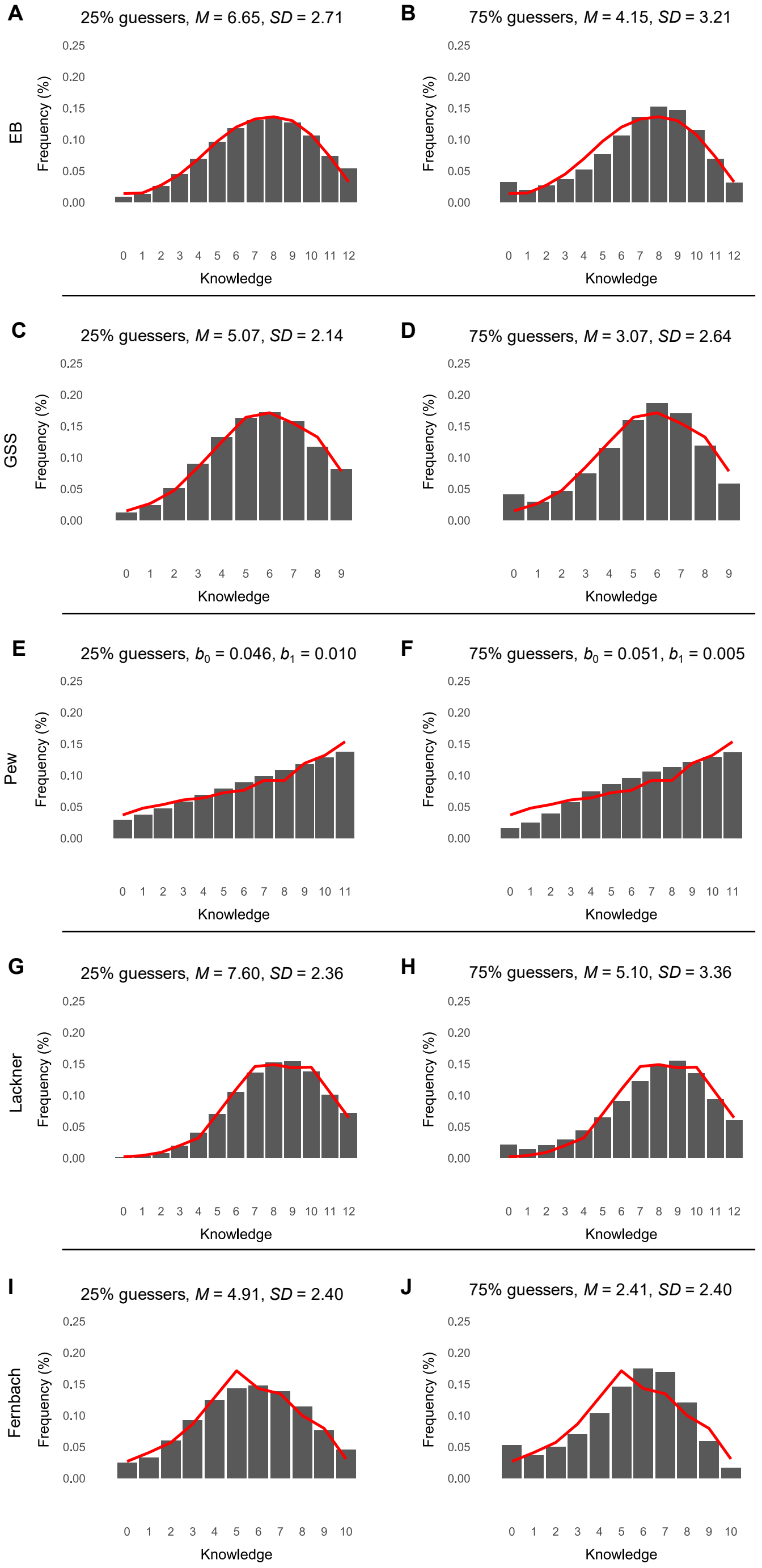

Supplementary Figure 9. Histogram of two different simulation models of the final distribution (i.e., the combination of guessers and agents who always say DK). Column on the left shows the distributions of the best fitting models (always found for 25% guessers, Supplementary Figure 8) for EB (A), GSS (C), Pew (E), Lackner (G), and Fernbach (I) with selected parameters on the top of each histogram (also from Supplementary Figure 8). The column on the right shows the histogram of the best fitting simulation models for the scenario of 75% guessers, to help visualize the poorer quality of fit of the best fitting scenarios, particularly at the extremes of the distribution, for EB (B), GSS (D), Pew (F), Lackner (H), and Fernbach (J). All distributions are approximated to a Normal, except for Pew, which is approximated to a one-sided Triangular distribution. Red line indicates the actual observed distribution of each dataset.

## Supplementary Figure 10. Comparison between the observed and the simulated proportions of incorrect answers

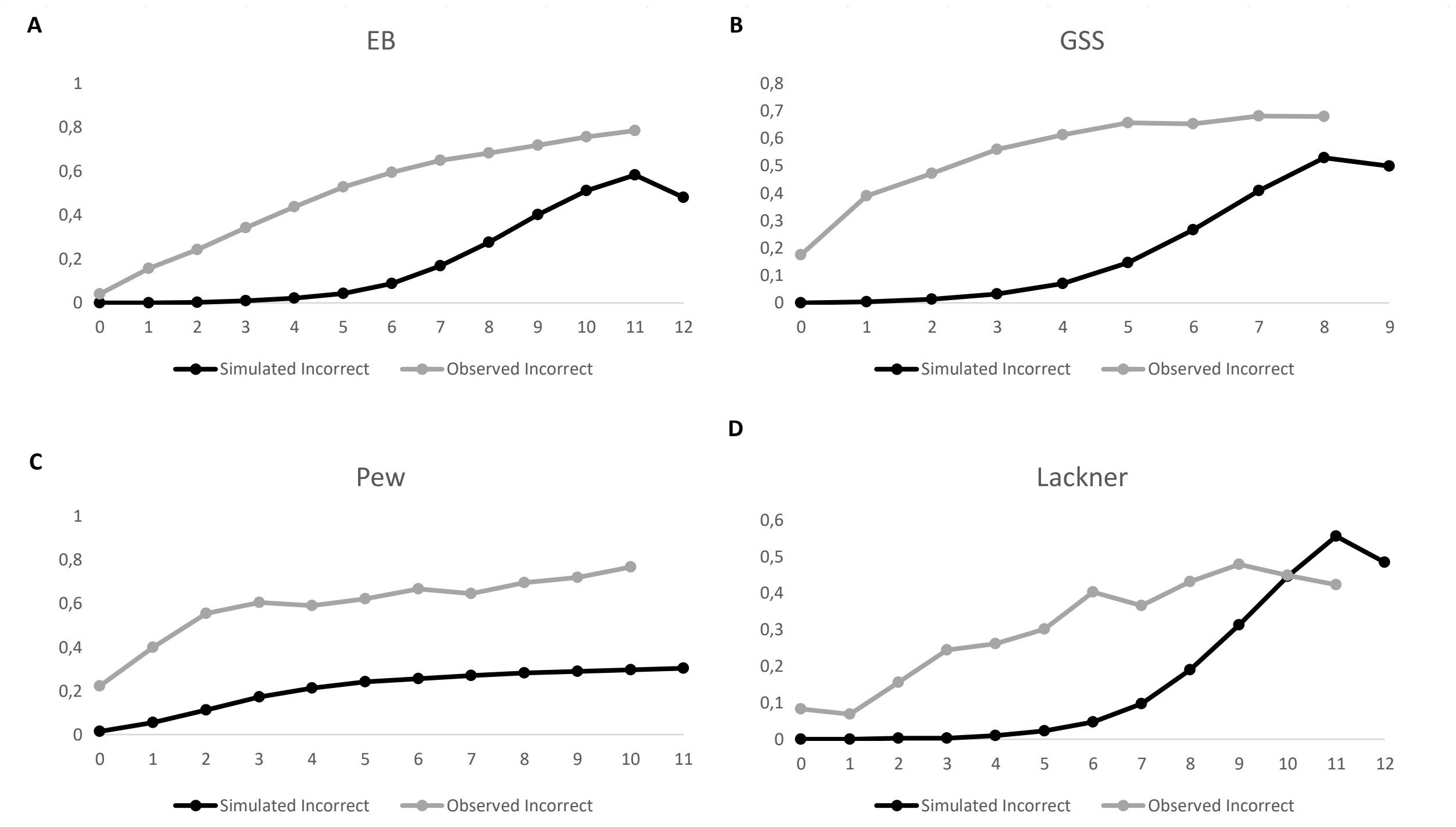

Supplementary Figure 10 – Comparison between the observed (grey lines) and the simulated (black lines) proportions of incorrect answers for each survey. Simulations were done assuming a similar baseline knowledge distribution for both guessers and non-guessers and using the parameters from Supplementary Figure 9 to define the percentage of guessers (25% in all plots).

## Supplementary Figure 11. Calibration Errors using the simulated distributions as baseline

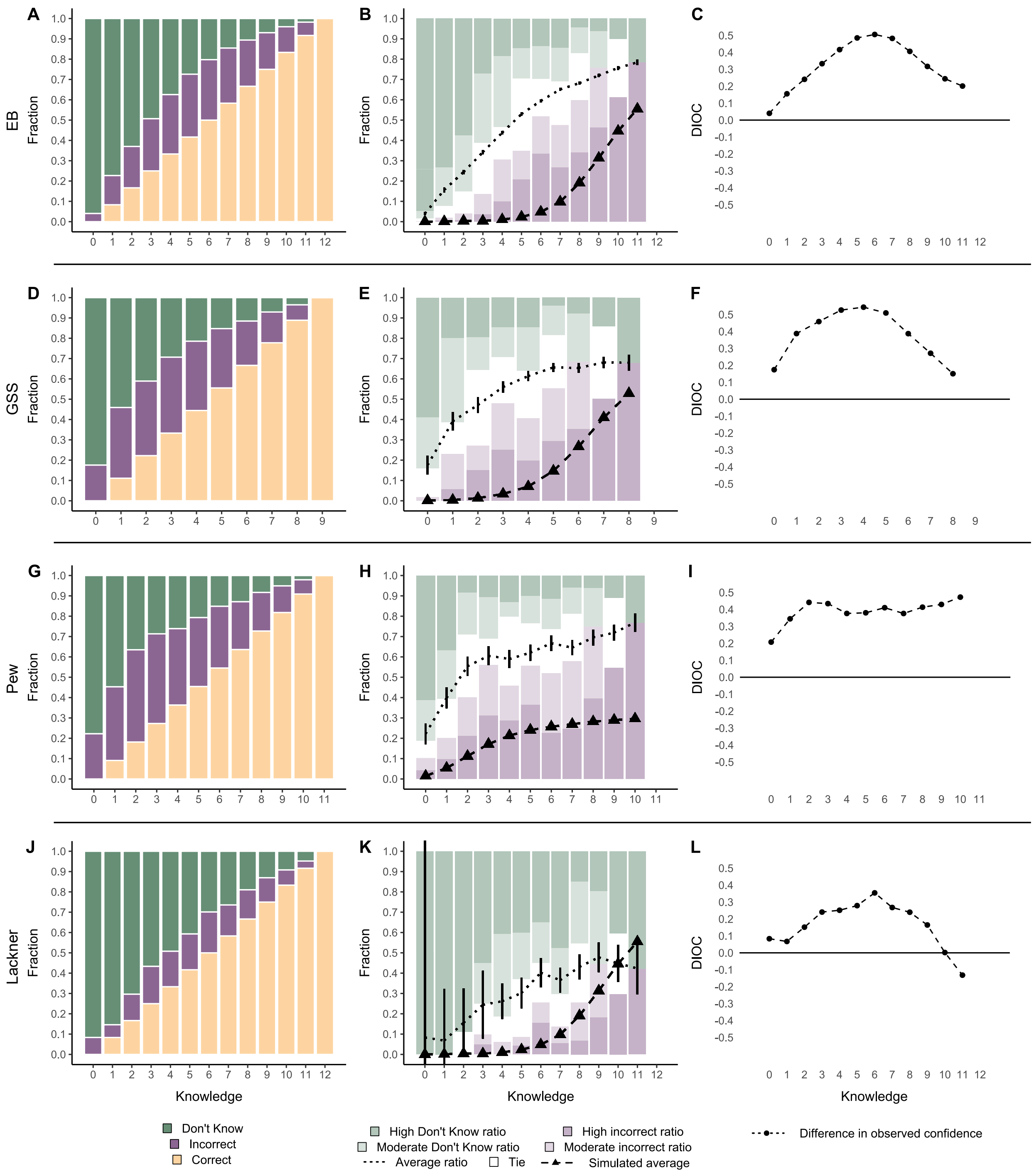

Supplementary Figure 11- Calibration errors using the simulated "incorrect" as baseline. Data from EB (**A, B, C**), GSS (**D, E, F**), Pew (**G, H, I**) and Lackner (**J, K, L**) surveys. In **A, D, G, J,** the subgroups of each column show the average fraction of respondents answering "Don't know" (green), incorrectly (purple) or correctly (yellow), per knowledge level (number of questions answered correctly), as in Figure 2 of the main manuscript. In **B, E, H, K, e**ach column shows the fractions of respondents according to knowledge level (proportion of "Don't Know" to incorrect answers by quintiles of normalized ratios):

Dotted black line shows the observed average normalized ratio (respondents are similarly likely to answer incorrectly or Don’t Know) and dashed black lines with arrows show the same ratio but for the simulated agents. C, F, I, L show the calibration error of the real (observed) respondents, using the simulated curves as baseline. In practical terms, the difference in observed confidence is calculated as the difference between the observed (dotted black line) and the simulated (dashed black line) curves. Bars correspond to 99% confidence intervals.

Supplementary Figure 12. EB Attitudinal data (complement to Figure 4).

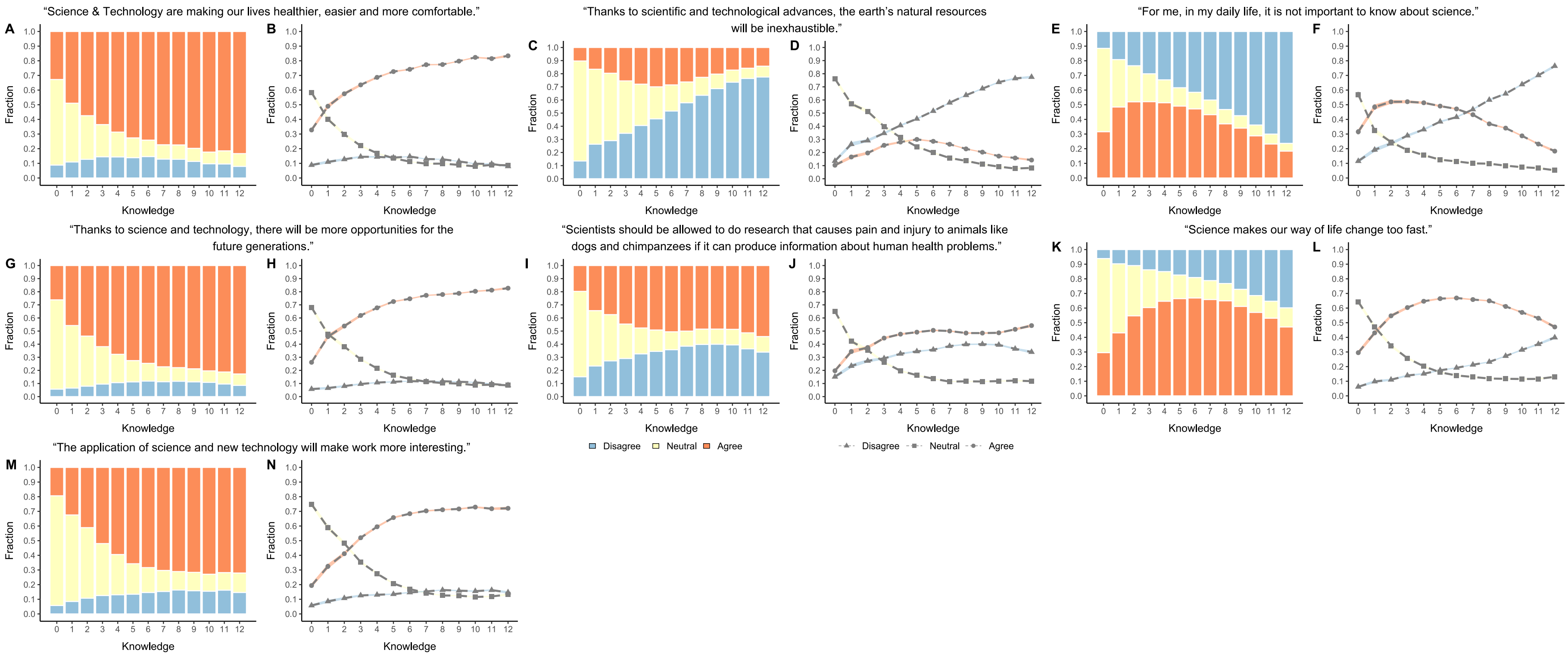

Supplementary Figure 12. EB Attitudinal data. (A,C,E,G,I,K,M) show stacked bar plots with fractions of Agree (orange), Neutral (yellow) and Disagree (red) answers in response to 7 EB attitude questions. Order of stacked bars is inverted in (E,K) as, in those two items, a negative attitude could be revealed by the Agree answer, while the reverse might be true for (A,G,M). (C) and (I) show a more nuanced response. Figures in (B,D,F,H,J,L,N) show the mean fractions across 34 EU territories with standard error of the mean.

Supplementary Figure 13. All attitudinal EB data.

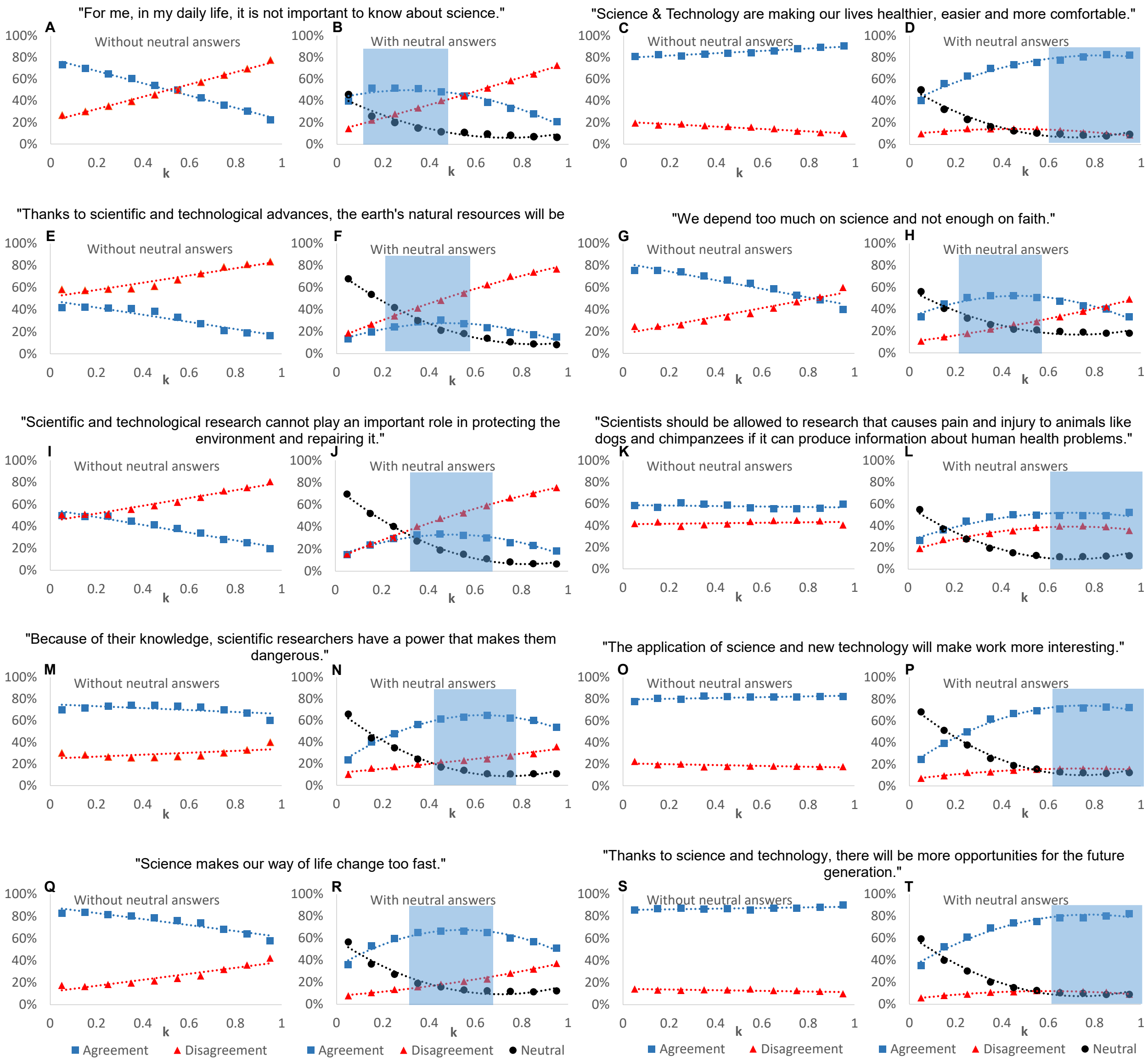

Supplementary Figure 13. All attitudinal EB data. Relative frequency of agreement (blue), disagreement (red), and neutral stance (black) for each knowledge category towards all EB attitude items. (A,C,E,G,I,K,M,O,Q,S) show relative frequency excluding neutral answers from the total, while (B,D,F,H,J,L,N,P,R,T) include all answers. Shaded areas in blue highlight the four consecutive knowledge bins with highest agreement in each attitude item. All the most negative attitudes (as perceived by the authors) are found in intermediate knowledge bins.Supplementary Figure 10. Spearman correlation matrix of attitude and knowledge variables (EB)

Supplementary Figure 14. Spearman correlation matrices of attitudes and knowledge.

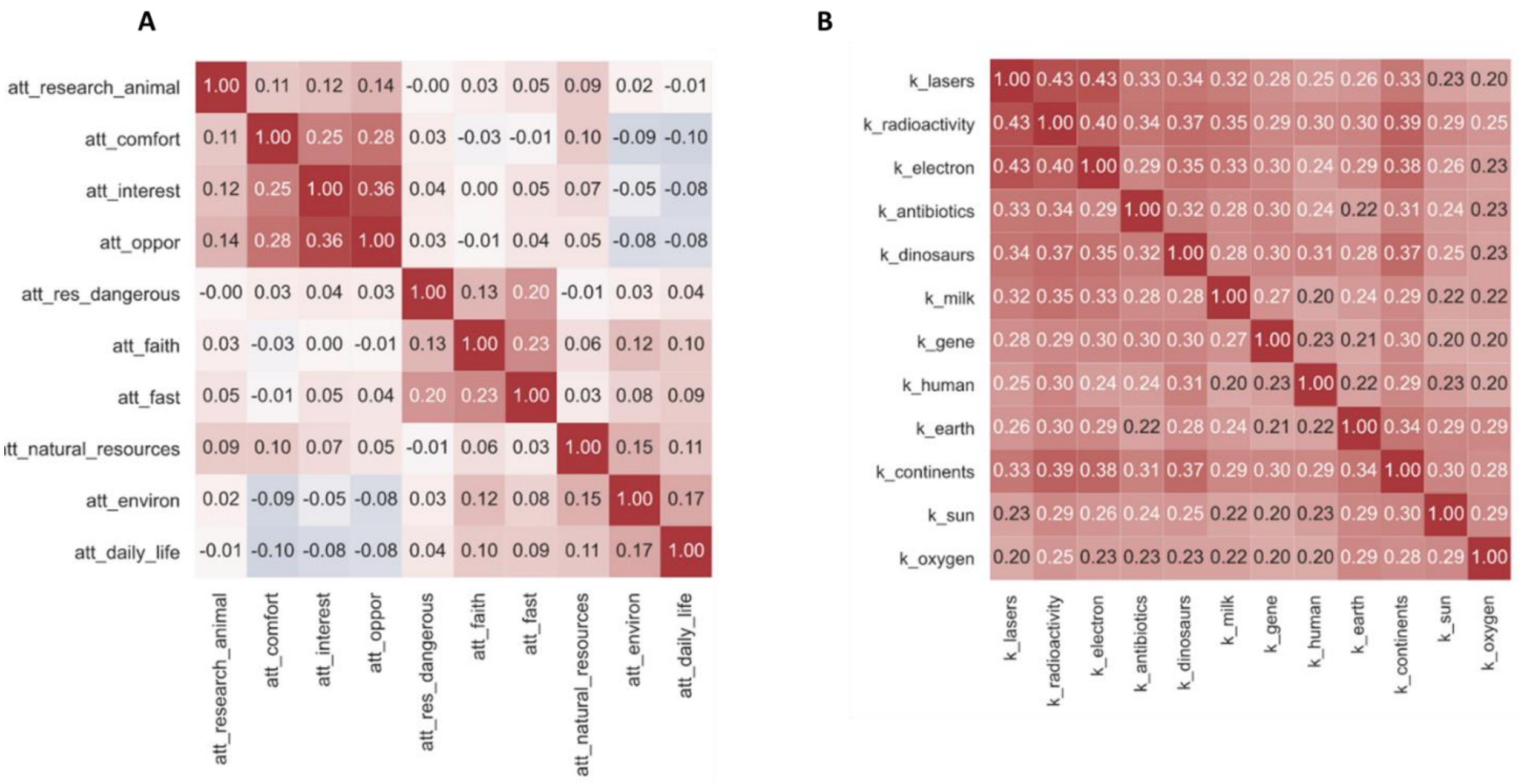

Supplementary Figure 14. EB data. Spearman correlation matrix of attitude (A) and knowledge (B) variables, ordered to show possible clusters. There is very little correlation between the attitudinal questions and medium between almost all knowledge questions.

Supplementary Figure 15. Proportion of variance for each PCA of the knowledge and attitude variables (EB).

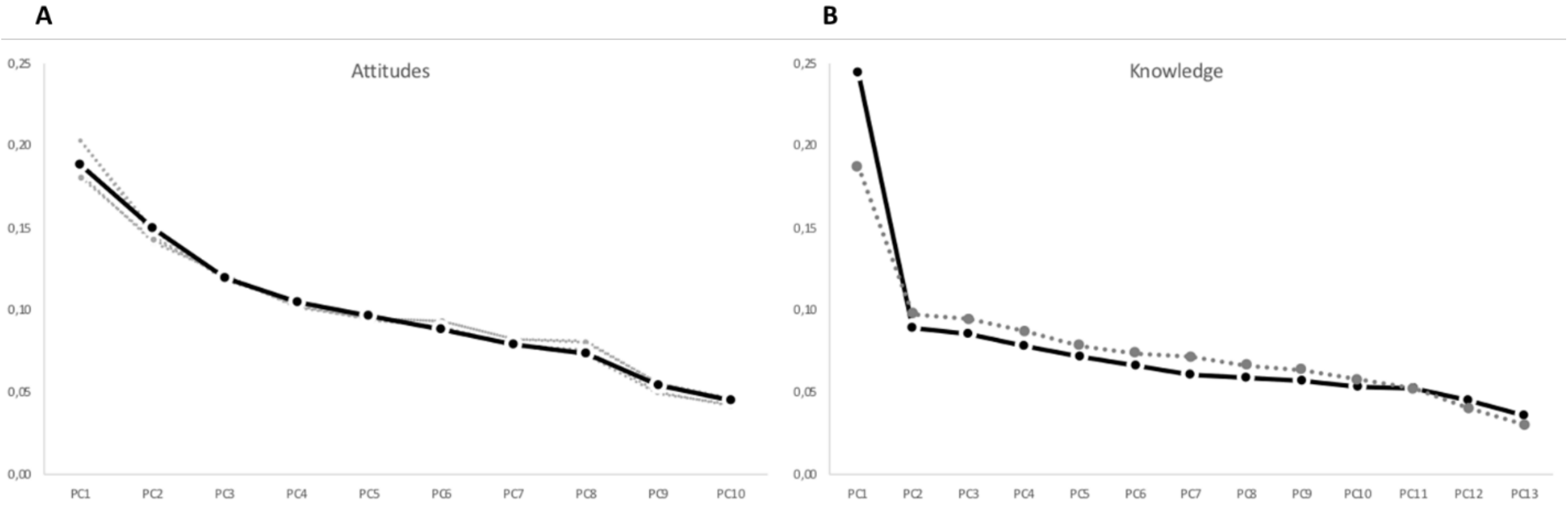

Supplementary Figure 15. EB data. Proportion of variance for each principal component resulting from the PCA ran on the knowledge and attitude variables. (A) Attitude variables PCA was performed by binning answers into three categories (positive, negative and neutral – solid black line) or treating the neutral answers separately (Don't know or no answer - grey dotted lines). The first principal components do not explain a large proportion of the total variance. (B) Knowledge variables PCA was performed either by aggregating "incorrect" and "Don't know" answers (solid black line) or treating them separately (grey dotted line). The first principal component accounts for a significantly larger part of the total variance and its coefficients all have the same sign.